 \documentclass[final,3p,times,twocolumn,10pt]{elsarticle}
\usepackage{booktabs}% Improve tabulars, enable \top-, \mid-, \bottomrule.
\usepackage{amsmath}
\usepackage{amsthm}
\usepackage{amssymb}
\usepackage{latexsym}
\usepackage{float}
\usepackage{graphicx}
\usepackage[font=normalsize,labelfont=bf]{caption}
\usepackage{subcaption}
\usepackage{epstopdf}
\usepackage[colorlinks=true,linkcolor=blue,citecolor=blue,urlcolor=blue]{hyperref}
\usepackage{siunitx}                 % Provides \si{} and \SI{}{}.
\usepackage[OMLmathsfit]{isomath} % Loads also fixmath.sty (upper case italic greek letters, definition of \upDelta etc).
\usepackage[flushleft]{threeparttable}
\usepackage{supertabular,booktabs,caption,fixltx2e}
\usepackage{longtable}
\usepackage{tabularx}                             % Specify total length of tabular.
  \newcolumntype{R}{>{\raggedleft\arraybackslash}X}  
  \newcolumntype{L}{>{\raggedright\arraybackslash}X} 
  \newcolumntype{C}{>{\centering\arraybackslash}X}
\renewcommand*{\vec}[1]{{\boldsymbol{#1}}}                       % Vector.
\newcommand*{\dd}{\mathrm{d}}                                    % Differential d for ``int dx''.

\newcommand{\specialcell}[2][c]{%
  \begin{tabular}[#1]{@{}c@{}}#2\end{tabular}}        
\journal{Fluid Phase Equilibria}

\begin{document}

\begin{frontmatter}

\title{Stabilized density gradient theory algorithm for modeling interfacial properties of pure and mixed systems}

\author[rvt]{Xiaoqun Mu}
\author[caam]{Florian Frank}
\author[focal]{Faruk O. Alpak}
\author[rvt]{Walter G. Chapman\corref{cor1}}
\ead{wgchap@rice.edu}
\cortext[cor1]{Corresponding author}

\address[rvt]{Department of Chemical and Biomolecular Engineering, Rice University}
\address[caam]{Department of Computational and Applied Mathematics, Rice University}
\address[focal]{Shell International Exploration and Production Inc.}

\begin{abstract}
  Density gradient theory (DGT) allows fast and accurate determination of surface tension and density profile through a phase interface. Several algorithms have been developed to apply this theory in practical calculations. While the conventional algorithm requires a reference substance of the system, a~modified ``stabilized density gradient theory''  (SDGT) algorithm is introduced in our work to solve DGT equations for multiphase pure and mixed systems. This algorithm makes it possible to calculate interfacial properties accurately at any domain size larger than the interface thickness without choosing a reference substance or assuming the functional form of the density profile. As part of DGT inputs, the perturbed chain statistical associating fluid theory (\mbox{PC-SAFT}) equation of state (EoS) was employed for the first time with the SDGT algorithm.  \mbox{PC-SAFT} has excellent performance in predicting liquid phase properties as well as phase behaviors. The SDGT algorithm with the PC-SAFT EoS was tested and compared with experimental data for several systems. Numerical stability analyses were also included in each calculation to verify the reliability of this approach for future applications.  
\end{abstract}

\begin{keyword}
%% keywords here, in the form: keyword \sep keyword
density gradient theory\sep stabilized algorithm\sep interfacial properties\sep surface tension\sep \mbox{PC-SAFT} EoS
%% PACS codes here, in the form: \PACS code \sep code

%% MSC codes here, in the form: \MSC code \sep code
%% or \MSC[2008] code \sep code (2000 is the default)

\end{keyword}

\end{frontmatter}

%% \linenumbers

%% main text
%\maketitle
%%%%%%%%%%%%%%%%%%%%%%%%%%%%%%%%%%%%%%%%%%%%%%%%%%%%%%%%%%%%%
\section{Introduction}
%%%%%%%%%%%%%%%%%%%%%%%%%%%%%%%%%%%%%%%%%%%%%%%%%%%%%%%%%%%%%
%Interfacial properties, including surface tension and interface density profile, play an important role in various industrial processes. For instance, 
In the petroleum industry, interfacial properties are of great interest since they affect various producing processes significantly such as gas injection and secondary oil recovery. An accurate and fast determination of these properties, for example surface tension and density profiles, is crucial in making appropriate business decisions and further to instruct industrial productions. Considering the fact that experimental measurements are costly and time-consuming, a~reliable theoretical method to predictively model interfacial properties is in high demand. Among many theoretical methods that have been developed so far, DGT is one of the most popular and successful method that has been applied in practical calculations.
\par
DGT was first proposed by Van der Waals \cite{rowlinson1979translation}. In his work, an interface area was described by a diffuse thin layer where a smooth density variation exists. The free energy is expressed by a function of the local density and its gradient. It was later reformulated by Cahn and Hillard \cite{cahn_free_1958} and DGT started to be widely studied subsequently. 
\par
Significant contributions to this theory were made by Carey et~al.~\cite{carey1979gradient} by reformulating the DGT equations to different differential equations in which a reference fluid is selected and the functions are solved accordingly. Later applications of DGT are largely based on Carey's reference fluid (RF) algorithm. Cornelisse et~al.~\cite{cornelisse_application_1993,cornelisse_non-classical_1996} compared the performance of DGT with the parachor method in several binary and ternary systems, and it was found that DGT is able to compute surface tension with a higher accuracy in various circumstances. Poser et~al.~\cite{poser_interfacial_1981} and Enders et al.~\cite{enders_calculation_1998}  applied this theory
to immiscible liquid--liquid phase interfaces. Teletzke et~al.~\cite{teletzke_gradient_1982} investigated wetting transitions using DGT. They modeled the physics of wetting qualitatively while additional experiments are still needed to determine it's quantitative accuracy.  In order to avoid the calculation of density profiles, Zuo and Stenby~\cite{zuo_linear_1996} developed linear gradient theory (LGT) by assuming that the density distributions of different components are independent of each other in a mixture. Miqueu et~al.~\cite{miqueu_modelling_2003,miqueu_modelling_2004,miqueu_modeling_2005} summarized the previous work and systematically developed DGT by generalizing its algorithm for multicomponent systems. 
\par
Although DGT gives the possibility to describe interfacial properties adequately, the lack of efficiency and robustness when solving the DGT equations bottlenecks its further development and not much progress \cite{zuo_linear_1996,kou2015efficient} has been made after Carey's reference fluid algorithm. In this paper, we developed a modified stabilized algorithm to improve the robustness of solving the DGT equations. Since the original DGT equations are rather sensitive to the chosen domain size, an~evolution term was introduced so that the ill-conditioned boundary value problem (BVP) becomes a~sequence of well-conditioned BVPs, each of which is solved using a~semi-implicit scheme in time by convex--concave splitting of the homogeneous free energy in a stable fashion.  Therefore, no reference fluid is needed in the calculation as opposed to the RF algorithm. The SDGT algorithm has several major advantages over the conventional RF algorithm and it shows great potential to operate with DGT for more complex systems.
\par
Cubic equations of state, including Van der Waals, Peng--Robinson, and Soave--Redlich--Kwong EoS, were widely used in DGT at early years \cite{carey1979gradient,cornelisse_application_1993,cornelisse_non-classical_1996,carey_semiempirical_1978,carey_semiempirical_1980,sahimi_thermodynamic_1985,sahimi_surface_1991,lin2007gradient,zuo_linear_1996,teletzke_gradient_1982}. While good results were obtained in vapor--liquid equilibrium system, significant errors were encountered in liquid--liquid equilibrium calculations, especially with the existence of associating fluids. In a series of papers \cite{chapman1988phase,chapman1989saft,chapman1990new}, 
Chapman et~al.\ introduced the statistical associating fluid theory (SAFT) EoS. Based on Wertheim's thermodynamic perturbation theory of first order \cite{wertheim1984fluids,wertheim1984fluids2,wertheim1986fluids,wertheim1986fluids2}, SAFT models the molecule by having spherical segments to form chains and counting the interactions among chain molecules. In comparison with other equations of state, SAFT demonstrates a much better performance in describing liquid densities and phase behavior \cite{vega201120years}. In our work, \mbox{PC-SAFT} \cite{gross2001perturbed,gross2002application} was employed to express the Helmholtz free energy as well as to conduct phase equilibrium calculations.
\par
This paper is organized as follows: In Section~\ref{sec:basictheory}, DGT and \mbox{PC-SAFT} EoS, including their basic theories and equations, are introduced.  Section \ref{sec:algorithm}  summarizes the existing DGT algorithms and their limitations. A~modified multiphase multicomponent SDGT algorithm is described afterwards. The performance of this algorithm was validated physically and numerically for several pure and mixture systems as presented in Section \ref{sec:results}.

%%%%%%%%%%%%%%%%%%%%%%%%%%%%%%%%%%%%%%%%%%%%%%%%%%%%%%%%%%%%%
\section{Theory}
\label{sec:basictheory}
%%%%%%%%%%%%%%%%%%%%%%%%%%%%%%%%%%%%%%%%%%%%%%%%%%%%%%%%%%%%%
\subsection{Density gradient theory}
\label{sec:basicdgt}
%%%%%%%%%%%%%%%%%%%%%%%%%%%%%%%%%%%%%%%%%%%%%%%%%%%%%%%%%%%%%
In DGT, the free energy~$A$ as functional of molar density fields~$\rho_i$, is derived as an~expansion about the free energy of a homogeneous fluid which can be expressed by an EoS, and the free energy of an inhomogeneous fluid which depends on the density gradient in that area (fourth and higher order gradient terms are neglected):
\begin{equation}\label{eq:originalF}
  A[\vec{\rho}] = \int_V \left[  a_0 (\vec{\rho})+\sum_{i,j=1}^{N} \frac{1}{2} v_{ij} \, \nabla \rho_i \cdot\nabla \rho_j \right]\,\dd V \,,
\end{equation}
where $N$~denotes the number of components in the system, $a_0$~the homogeneous Helmholtz free energy density, which is given by a~bulk EoS, and $v_{ij}$~the so-called influence parameter.
 In most cases, the density-dependence of the influence parameter is neglected so that $\partial v_{jk}/\partial \rho_i=0$.
The vector of molar densities~$\vec{\rho}$ in arguments indicate the dependency of all molar densities~$\rho_i$. 
\par
  In an open, isothermal system with no external fields, the grand potential~$\Omega$ can be expressed as:
  \begin{equation*}
    \Omega[\vec{\rho}] = A[\vec{\rho}] - \int_V \sum_{i=1}^N \rho_i\, \mu_{i,\mathrm{bulk}}   \,\dd V\,,
  \end{equation*}
  where $\mu_{i,\mathrm{bulk}}$ is the (constant) bulk chemical potential of component~$i$.
  When the system has reached a~stationary state, the grand potential is minimized, i.\,e.~the functional derivatives of~$\Omega$ vanish:
  \begin{multline}\label{eq:ELequation}
     \frac{\delta \Omega[\vec{\rho}]}{\delta \rho_i}
     =
     \frac{\partial a_\mathrm{0}(\vec{\rho})}{\partial \rho_i} 
      -  \mu_{i,\mathrm{bulk}} 
      -\sum_{j=1}^N v_{ij} \, \nabla \cdot\nabla \rho_j 
     =
     0\,,
  \end{multline}
  for $i=1,\ldots,N$, i.\,e.,~the molar densities of the system must satisfy the Euler--Lagrange equation at stationary state.  Here, we assumed~$v_{ij} = v_{ji}$ (cf.~Eqn.~\eqref{eq:mixingrule:vij}).
  \par
  In the case of a~planar interface, the density only varies in $z$-direction, and Eqn.~\eqref{eq:ELequation} simplifies to
  \begin{subequations}\label{eq:simplifiedEL1:BVP}
  \begin{equation}\label{eq:simplifiedEL1}
    \mu_i(\vec{\rho})-\mu_{i,\mathrm{bulk}} = \sum_{j=1}^{N} v_{ij} \frac{\dd^2 \rho_j}{\dd z^2}\,,
  \end{equation}
  where the (homogeneous) chemical potential of component~$i$ is given by~$\mu_i(\vec{\rho}) = {\partial a_\mathrm{0}(\vec{\rho})}/{\partial \rho_i}$.
  Solving Eqn.~\eqref{eq:simplifiedEL1} yields the density distributions~$\rho_i(z)$ across the planar interface region.   
  It is subjected to the following boundary conditions:
  \begin{equation}\label{eq:simplifiedEL1:bc}
    \rho_i(0) = \rho_{i,\mathrm{A}} \,, \qquad  \rho_i(D) = \rho_{i,\mathrm{B}} \,, 
  \end{equation}
  \end{subequations}
  for $i=1,\ldots, N$,
  where $\rho_{i,\mathrm{A}}$ and $\rho_{\mathrm{i,B}}$ are (constant) bulk densities of component~$i$ in phase~A and~B, respectively. 
  The symbol~$D$ denotes any distance that is greater than the interface thickness~$L$ in order to ensure that the domain boundaries are located in bulk phases.
  Once the density profiles are determined, the surface tension~$\sigma$ is evaluated by 
  \begin{equation}\label{eq:sigma1}
    \sigma=\int_{-\infty}^{\infty} \sum_{i=1}^N \sum_{j=1}^N v_{ij} \frac{\dd\rho_i}{\dd z}\frac{\dd\rho_j}{\dd z} \,\dd z\,.
  \end{equation}
  \par
  Values of the pure component influence parameters~$v_i$ (cf.~Eqn.~\eqref{eq:mixingrule:vij}) are obtained by fitting with experimentally measured surface tension at fixed temperature.  Influence parameters used in this paper are listed in Table~\ref{tab:PCSAFTinflupara}. 
  \begin{table}
      \caption{Influence parameters~$v_i$ for pure components (\mbox{PC-SAFT}).}
      \label{tab:PCSAFTinflupara}
      \begin{tabularx}{\linewidth}{@{}LCCC@{}}
        \toprule
        Component              &  \specialcell{$T$\\$[\si{K}]$}&  \specialcell{$\sigma$ \\$[\si{mN/m}]$} & \specialcell{$v_i\cdot10^{20} $\\$[\si{J\,m^5/mol^2}]$}\\
        \midrule
        Methane  &  104.50 &14.36 \cite{Baidakov1982}  &1.995    \\
        Propane    &   332.92 & 3.09  \cite{baidakov1985surface}&10.460   \\
        n-Pentane &  249.34&20.42 \cite{Grigoryev1992} &24.779  \\
        n-Hexane &  244.81 &23.66 \cite{Grigoryev1992}  &35.575  \\
        Toluene &332.15 &23.88 \cite{kalbassi1988surface}  &32.152      \\  
        \bottomrule    
      \end{tabularx}
    
  \end{table}
  
  \subsection {\mbox{PC-SAFT} EoS}
  The SAFT EoS was originally developed by Chapman~et~al.\ by a series of papers~\cite{chapman1988phase,chapman1989saft,chapman1990new} using an extension of Wertheim's thermodynamic perturbation theory of first order~(TPT1). In the SAFT framework, a system in which only hard sphere segments exist is defined as a reference fluid. Based on this, mixtures of polyatomic associating molecules are modeled by adding perturbations of the association interactions to the reference fluid as well as the chain formation contributions by assuming infinitely strong association attractions between hard spheres. 
  \par
  Later, the contribution from long range attractions (dispersion) in SAFT was revisited and improved by Gross and Sadowski who developed the \mbox{PC-SAFT} EoS \cite{gross2001perturbed,gross2002application}. In \mbox{PC-SAFT},  a~system with only hard chain repulsion force is defined as a reference fluid instead. The perturbation theory of Baker and Henderson \cite{barker1967perturbation} was introduced to the reference system. The homogeneous Helmholtz free energy of \mbox{PC-SAFT} is expressed as:
  \begin{equation}\label{eq:homogeneousFreeEnergy}
    A_0 = A_0^\mathrm{ideal}+A_0^\mathrm{hs}+A_0^\mathrm{hc}+A_0^\mathrm{disp}+A_0^\mathrm{assoc}~,
  \end{equation}
  where $A^\mathrm{ideal}$ is the ideal gas Helmholtz free energy known from thermodynamics, $A_0^\mathrm{hs}$ and $A_0^\mathrm{hc}$ are the Helmholtz free energy due to the hard spheres and the formation of hard chains respectively, $A_0^\mathrm{disp}$ is the Helmholtz free energy of dispersion attraction and  $A_0^\mathrm{assoc}$ accounts for the associating energy between molecules. 
  \par
  For non-associating substances, three \mbox{PC-SAFT} parameters are required: $m_i$, the effective number of segments within molecule which represents the chain length; $\sigma_i$, the diameter of each segment; and $\epsilon_i$, the depth of pair potential energy between same segment. Another two parameters are necessary for substances with association sites: $\epsilon^{\mathrm{A}_i\mathrm{B}_i}$, the association energy of interaction and $\kappa^{\mathrm{A}_i\mathrm{B}_i}$ the effective volume of interaction  between site~A and site~B on molecule~$i$. 
  \par
  When applied in mixtures, parameters of binary  component combinations are calculated by the following mixing rules:
  \begin{align*}
      \sigma_{ij}&=\frac{1}{2}(\sigma_i+\sigma_j)\,,\\
      \epsilon_{ij}&=(1-k_{ij})\sqrt{\epsilon_i\epsilon_j}\,,
  \end{align*}
  where $k_{ij}$ is the binary interaction parameter. If associating interactions exist, the cross-association parameters can be determined using the mixing rule suggested by Wolbach and Sandler~\cite{wolbach1998using}:
  \begin{align*}
      \epsilon^{\mathrm{A}_i\mathrm{B}_j}&=\frac{1}{2}(\epsilon^{\mathrm{A}_i\mathrm{B}_i}+\epsilon^{\mathrm{A}_j\mathrm{B}_j})\,,\\
      \kappa^{\mathrm{A}_i\mathrm{B}_j}&=\sqrt{\kappa^{\mathrm{A}_i\mathrm{B}_i}\kappa^{\mathrm{A}_j\mathrm{B}_j}}\left(\frac{\sqrt{\sigma_{ii}\sigma_{jj}}}{\frac{1}{2}(\sigma_{ii}+\sigma_{jj})}\right)^3\,.
  \end{align*}
  \mbox{PC-SAFT} parameters used in this paper are listed in Table~\ref{tab:PCSAFTpara1}. For more details about the derivations and parameters of \mbox{PC-SAFT}, one can refer to the original SAFT \cite{chapman1988phase,chapman1989saft,chapman1990new} and \mbox{PC-SAFT} papers \cite{gross2001perturbed,gross2002application}.
   \begin{table}\small
   	\caption{\mbox{PC-SAFT} parameters of non-associating pure component.} 
   	\label{tab:PCSAFTpara1}
   	\begin{tabularx}{\linewidth}{@{}lCCcC@{}}
   		\toprule
   		Component& \specialcell{$M_i$ \\$[\si{g/mol}]$} & \specialcell{$m_i$\\$[1]$} & \specialcell{$\sigma_i$ \\$[\AA{}]$} &\specialcell{ $\epsilon_i/k_B$ \\ $[\si{K}]$} \\
   		\midrule
   		Methane &  16.043  &  1.0000 &  3.7039 &  150.03   \\
   		Propane &  44.096   & 1.6069   &  3.5206  &  191.42  \\
   		n-Pentane &  72.146   & 2.6896   &  3.7729  &  231.2 \\
   		n-Hexane &  86.177   & 3.0576   &  3.7983  &  236.77 \\
   		Toluene&  92.141 &  2.8149  &  3.7169 &  285.69  \\  
   		\bottomrule
   	\end{tabularx}
   \end{table}
  %%%%%%%%%%%%%%%%%%%%%%%%%%%%%%%%%%%%%%%%%%%%%%%%%%%%%%%%%%%%%
  \section{Challenges and algorithms}
  \label{sec:algorithm}
  %%%%%%%%%%%%%%%%%%%%%%%%%%%%%%%%%%%%%%%%%%%%%%%%%%%%%%%%%%%%%
  Derived by minimizing the grand potential energy ~$\Omega$, the BVP~\eqref{eq:simplifiedEL1:BVP} is solved to obtain the equilibrium density profile, and the surface tension is calculated accordingly.
  However, certain numerical challenges exist in the solving process of this BVP:
  Although theoretically the domain size~$D$ in the boundary conditions can be any value that is greater than the interface thickness~$L$, a~stable convergence in the solving process will happen only if a close estimation of~$D$ to~$L$ is given. This is because Eqn.~\eqref{eq:simplifiedEL1:BVP} is rather sensitive to the boundary values and becomes ill-conditioned with an~overestimated or underestimated value of~$D$. The numerical nonlinear solver will have severe stability issues and convergence failure occurs easily. Nevertheless, finding an~adequate estimation of the interface thickness for an unknown system is challenging, which makes this BVP fairly difficult to solve.
  
  \subsection{Established algorithms}
   Different algorithms have been developed to tackle the stability issues encountered in solving DGT equations, such as the LGT algorithm~\cite{zuo_linear_1996} and the RF algorithm~\cite{carey1979gradient,sahimi_thermodynamic_1985}. 
   
  \subsubsection{Linear gradient theory}
   The LGT algorithm simply assumes that intermolecular interactions have no impact on interface density distributions, and thus the density profiles of each component are calculated independently in a mixture. These assumptions simplify the DGT model and makes the calculation faster, but it loses most of the interface physics at the same time. Therefore, the LGT algorithm is not recommended.
  
  \subsubsection{Reference fluid algorithm}
  The reference fluid (RF) algorithm, meanwhile, is the most widely used algorithm so far. According to this algorithm, one component  $\rho_\mathrm{ref}$ is selected as a reference fluid in the system and by certain manipulations \cite{sahimi_thermodynamic_1985}, the original differential equations of  $\rho_i(z)$ defined in an unknown domain~$[0,D]$ are transformed to a~problem of $\rho_i(\rho_\mathrm{ref})$ for $i \neq \mathrm{ref}$ in the~known domain~$[\rho_\mathrm{ref,A},\rho_\mathrm{ref,B}]$ with boundary conditions
  \begin{equation*}
      \rho_i(\rho_\mathrm{ref,A}) = \rho_{i,\mathrm{A}}\, , \qquad  \rho_i(\rho_\mathrm{ref,B}) = \rho_{i,\mathrm{B}}\, .
  \end{equation*}
  Here, $\rho_\mathrm{ref,A}$ and $\rho_\mathrm{ref,B}$ are determined directly by phase equilibrium calculations---no estimation of interface thickness is needed. Solving the RF DGT yields the density dependence of each substance to the reference fluid, i.\,e. $\rho_i(\rho_\mathrm{ref})$. With these results, the density profile and surface tension are calculated by:
  \begin{align*}
    z&=z_0 +\int_{\rho_\mathrm{ref,A}}^{\rho_\mathrm{ref}} \sqrt{\frac{C}{2(a_0-\sum_i \rho_i \mu_{i,\mathrm{bulk}}+P_0)}} \,\dd \rho_\mathrm{ref}\,,
\\
    \sigma&=\int_{\rho_\mathrm{ref,A}}^{\rho_\mathrm{ref,B}} \sqrt{2C(a_0-\sum_i \rho_i \mu_{i,\mathrm{bulk}}+P_0)  } \,\dd\rho_\mathrm{ref}\,,
  \end{align*}
  where $P_0$ is the bulk pressure and 
  \begin{equation*}
    C=\sum_{i=1}^N\sum_{j=1}^Nv_{ij}\frac{\dd \rho_i}{\dd \rho_\mathrm{ref}}\frac{\dd \rho_j}{\dd \rho_\mathrm{ref}}\, .
  \end{equation*}
  \subsubsection{Limitations}
  By reformulating the DGT equations, the RF algorithm makes the solving process numerically straightforward and it has been successfully applied to calculate interfacial properties in many vapor--liquid and liquid--liquid equilibrium systems~\cite{carey1979gradient}--\cite{lin2007gradient}. However, this algorithm has several drawbacks that limit the application of DGT to a wider range of systems:
  \par
  First, no general strategy of selecting the reference fluid is available in this algorithm. Although boundary conditions can be provided by phase equilibrium calculations, a~suitable reference fluid needs to be chosen before starting the calculation. The main requirement of the reference fluid is that its density must be a monotonic function of the distance~$z$ across the interface. If the monotonicity of a preselected reference fluid changes in the interface area, it must be switched to a~different component according to the density function behavior, and this procedure will be repeated if the new reference fluid becomes non-monotonic again. This is a time-consuming process. More importantly, failure to select the suitable reference fluid will lead to numerical errors in the solving process and some interfacial phenomena like surface density accumulations cannot be described correctly. In most cases, the determination of the reference fluid is based on experience, introducing a factor of uncertainty to the calculations. 
  \par
  Second, the RF algorithm is no longer valid when any additional term is added to the DGT functional. As we mentioned before, the BVP solved in the RF algorithm is transformed from the original DGT equations~\eqref{eq:originalF}--\eqref{eq:simplifiedEL1:BVP} by certain manipulations (more details about this manipulation process can be found from Carey et~al.~\cite{carey1979gradient} or Sahimi et~al.~\cite{sahimi_thermodynamic_1985}). Noticeably, the prerequisite of this manipulation process is that the original DGT functional form must be strictly followed. In other words, the RF algorithm, if applicable, works exclusively for this specific DGT functional form. Any extensions of DGT functions are not allowed or the manipulation process won't work. This will prevent the future application of DGT to systems like polymer or colloidal mixtures, in which molecules are considered to have more complex structures that require additional energy terms.
  
  \subsection{Stabilized DGT algorithm}
  \subsubsection{Modified DGT equations}
  Since there exists the dual challenges of selecting a~suitable reference fluid as well as extending DGT functional forms in the RF algorithm, a~novel and effective SDGT algorithm was developed for DGT equations in this paper.
  \par
  We revisited the original DGT model. As was discussed before, solving the BVP~\eqref{eq:simplifiedEL1:BVP} requires a~good estimation of the interface thickness~$L$ or severe numerical issues will be encountered. In order to avoid this, an evolution term $\partial \rho_i/\partial s$ was added to Eqn.~\eqref{eq:simplifiedEL1} such that it becomes a~time-dependent partial differential equation. This idea was also used by Qiao and Sun~\cite{qiao2014two}, where they applied DGT with Peng--Robinson EoS in a~single component vapor--liquid equilibrium system.  In our work, we generalize this algorithm to a~multiphase multicomponent system with \mbox{PC-SAFT} EoS.
  \par
  In a~system with $N$~components, the SDGT algorithm has the following form:
  \begin{subequations}\label{eq:SDGT} % SDGT = stabilized DGT
    \begin{equation}\label{eq:SDGT:PDE}
    \frac{\partial \rho_{i}}{\partial s} + \mu_i(\vec{\rho}) - \mu_{i,\mathrm{bulk}} = \sum_{j=1}^N v_{ij} \frac{\partial^2 \rho_j}{\partial z^2}\;,
  \end{equation}
  for $i=1,\ldots,N$, where $\rho_i = \rho_i(s,z)$ is now to be considered a~function also of~$s$, which may be interpreted as time variable.
  This equation is subjected to the boundary conditions
   \begin{equation}\label{eq:SDGT:BC}
    \rho_i(s, 0) = \rho_{i,\mathrm{A}} \,, \qquad  \rho_i(s, D) = \rho_{i,\mathrm{B}} \,, 
  \end{equation}
  for all time points~$s$.  In Eqn.~(\ref{eq:SDGT:PDE}), the time derivative~$ \partial \rho_{i} / \partial s$ serves as a~stabilizing term to ensure a stable convergence. These BVP functions do not preserve mass (open system) and thus give a lot of freedom in choosing initial data. An~easy and good initial condition for this system of differential equations is the~linear density distribution of each component across the domain interpolating the boundary values of Eqn.~\eqref{eq:SDGT:BC}:
  \begin{equation}\label{eq:SDGT:INIT}
      \rho_i (0,z) = \rho_{i,\mathrm{A}} + \frac{\rho_{i,\mathrm{B}} - \rho_{i,\mathrm{A}}}{D}\,z \,.
  \end{equation}
  \end{subequations}
  The system of time-dependent BVPs~\eqref{eq:SDGT} is solved with a~time marching scheme until a~stationary state is reached.
  In fact, the SDGT algorithm is quite robust with regards to initial conditions---even when an~unfavorable density distribution estimation is chosen.  This attribute will be further discussed in Section~\ref{sec:results}.
  \subsubsection{Time Discretization}   \label{sec:initialguess}
  In order to apply the new algorithm more efficiently, a~convex--concave splitting of the non-linear energy was followed, cf.~\cite{ElliottStuart1993,eyre1998unconditionally}. 
  The convex part of the homogeneous free energy~$A_0$ in~Eqn.~\eqref{eq:homogeneousFreeEnergy} is treated time-implicitly using the backward Euler method, while the concave part is treated time-explicitly by a~forward Euler method. This splitting scheme makes the time discretization unconditionally stable, i.\,e.~there is no restriction in time step size (however, large time steps imply a~large condition number in the linear system that has to be solved in every Newton iteration cf.~Section~\ref{sec:spacedis}). It also ensures a monotonic dissipation of free energy with respect of time (Fig.~\ref{fig:gphexane},~\ref{fig:gp_methane_propane}).
  \par
  It is simple to prove that the ideal gas Helmholtz free energy~$A_0^\mathrm{id}$ is a~convex function with respect to~$\rho_i$. In the excess Helmholtz free energy, the repulsion force between molecules, as justified by Qiao and Sun in their work~\cite{qiao2014two}, must result in a~convex contribution while the attraction force should have a concave contribution so as to have phase splitting occur.  In \mbox{PC-SAFT}, we have the hard sphere~$A_0^\mathrm{hs}$ and hard chain~$A_0^\mathrm{hc}$ contributions to the excess Helmholtz free energy as a~result of repulsion forces, and association~$A_0^\mathrm{assoc}$ and dispersion~$A_0^\mathrm{disp}$ contributions as a result of attraction forces, cf.~Eqn.~\eqref{eq:homogeneousFreeEnergy}:
  \begin{align*}
      A_0^\mathrm{convex}   &= A_0^\mathrm{id}+A_0^\mathrm{hs}+A_0^\mathrm{hc} \,,\\
      A_0^\mathrm{concave}  &= A_0^\mathrm{disp}+A_0^\mathrm{assoc}\,.
  \end{align*}
  \par 
  Discretizing Eqn.~(\ref{eq:SDGT:PDE}) in time while applying the convex--concave splitting yields:
  \begin{multline*}
      \frac{\rho_i^{n+1} - \rho_i^n}{\Delta s} + \mu^\mathrm{convex}_{i}  (\vec{\rho}^{n+1}) + \mu^\mathrm{concave}_{i}(\vec{\rho}^n)
      \\
      =\mu_{i,\mathrm{bulk}} + \sum_{j=1}^N v_{ij} \frac{\dd^2 \rho_j^{n+1}}{\dd z^2}\,,
  \end{multline*} 
where $\mu^\mathrm{convex}_i(\vec{\rho}) = {\partial a^\mathrm{convex}_\mathrm{0}(\vec{\rho})}/{\partial \rho_i}$ and $\mu^\mathrm{concave}_i(\vec{\rho}) = {\partial a^\mathrm{concave}_\mathrm{0}(\vec{\rho})}/{\partial \rho_i}$.
\subsubsection{Space discretization and boundary conditions} \label{sec:spacedis}
In each time step, we have a~system of $N$~nonlinear equations that are solved by finite difference method using Newton's iteration.  Stepping forward in time, the system will evolve to an~equilibrium state, which is
reached when the stopping criteria
  \begin{equation}
    \sum_{i=1}^N\sum_{k=1}^{M}~\lvert \rho_{i,k}^{n+1}-\rho_{i,k}^{n} \rvert  < \varepsilon\,
  \end{equation}
  is satisfied, where $M$ is the number of grid points in space, $0<\varepsilon\ll 1$ a~tolerance, and $\rho_{i,k}^{n+1}$ the density of component~$i$ in position~$k$ at time step~$n+1$.  The obtained equilibrium density distribution is an~approximation of the solution of the original DGT model~\eqref{eq:simplifiedEL1:BVP}. Surface tension can be computed via Eqn.~(\ref{eq:sigma1}).
 \begin{figure}[ht!]
 	{
 		\centering
 		\begin{subfigure}[t]{0.475\linewidth}
 			\includegraphics[width=\linewidth]{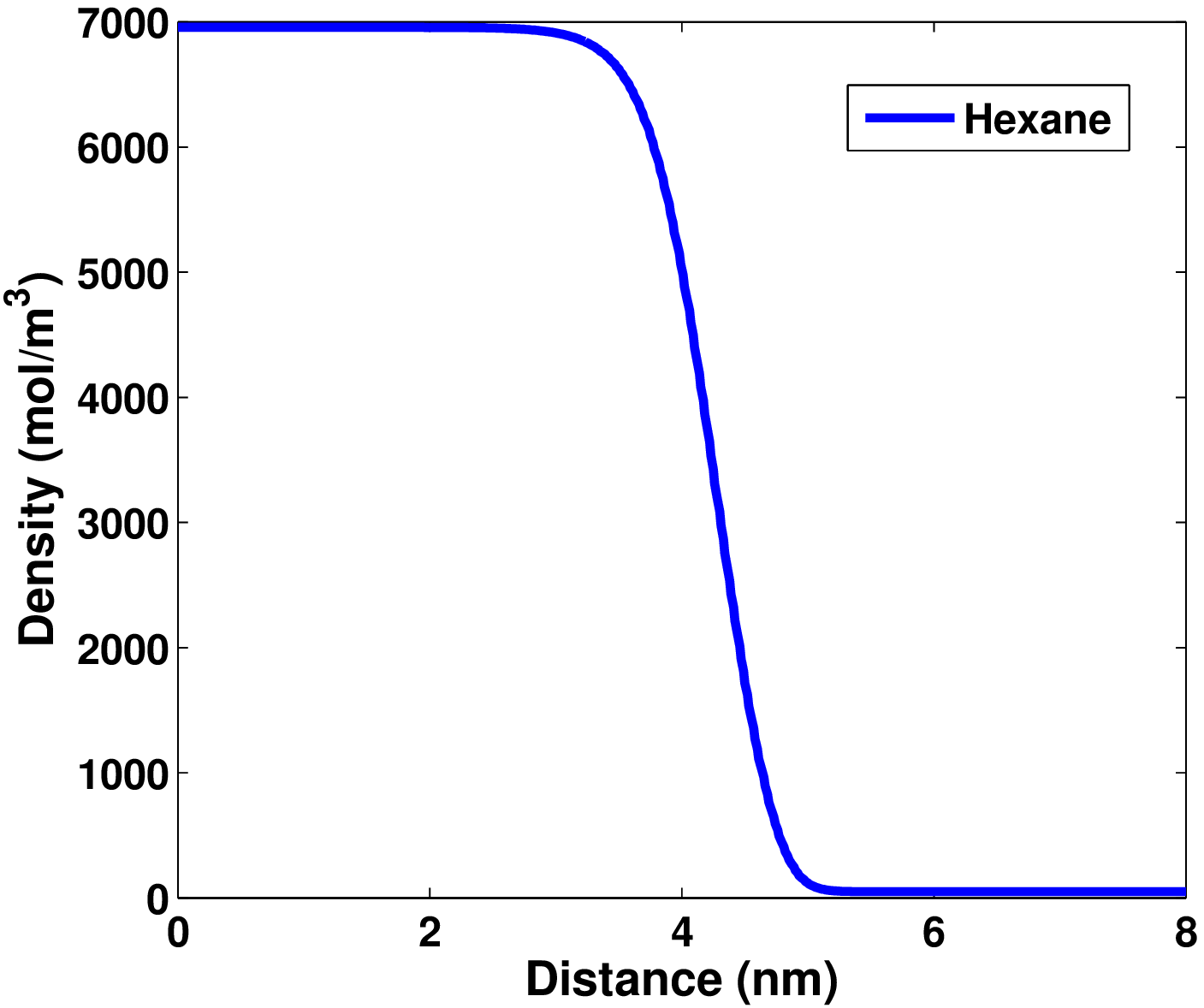}
 			\caption{Density profile of hexane (352.49~K) after 10~time steps on $D=8~\mathrm{nm}$. Diffuse interface occurs in the middle of the domain.} 
 		\end{subfigure}
 		\hfill
 		\begin{subfigure}[t]{0.475\linewidth}
 			\includegraphics[width=\linewidth]{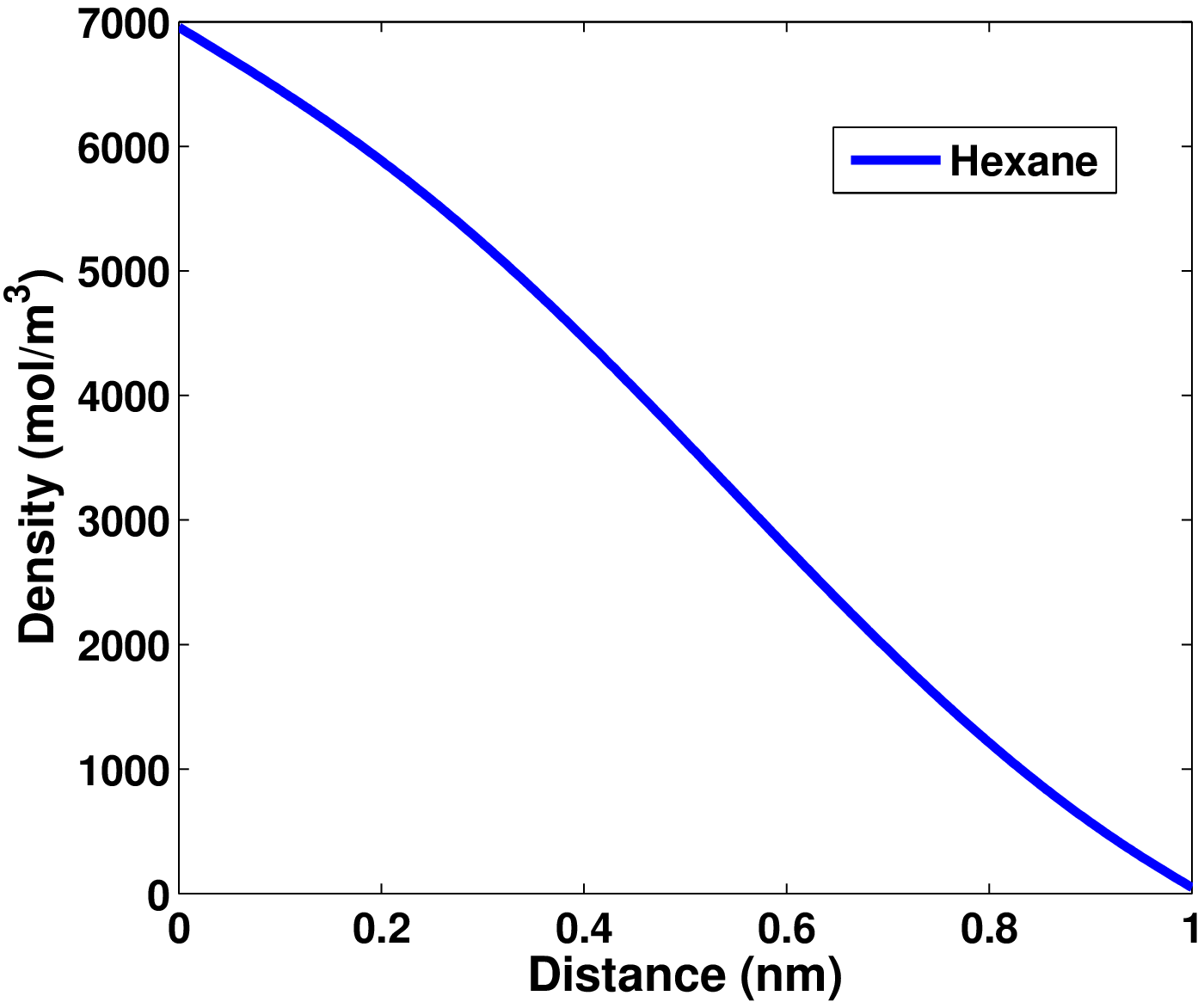}
 			\caption{Density profile of hexane (352.49~K) after 10~time steps on $D=1~\mathrm{nm}$. Diffuse interface occurs right next to boundaries.} 
 		\end{subfigure}
 	}
 	\caption{A~domain size $D$ that is wider than the interface thickness $L$ is needed for a correct convergence as shown in (a). If an underestimation of $D$ is given, it can be detected easily from the shape of the density profile as shown in (b).}
 	\label{fig:bcdiscussion}
 \end{figure}
 
    \begin{figure*}[ht!]
      \centering
      \begin{subfigure}[t]{0.245\linewidth}
        \includegraphics[width=\linewidth]{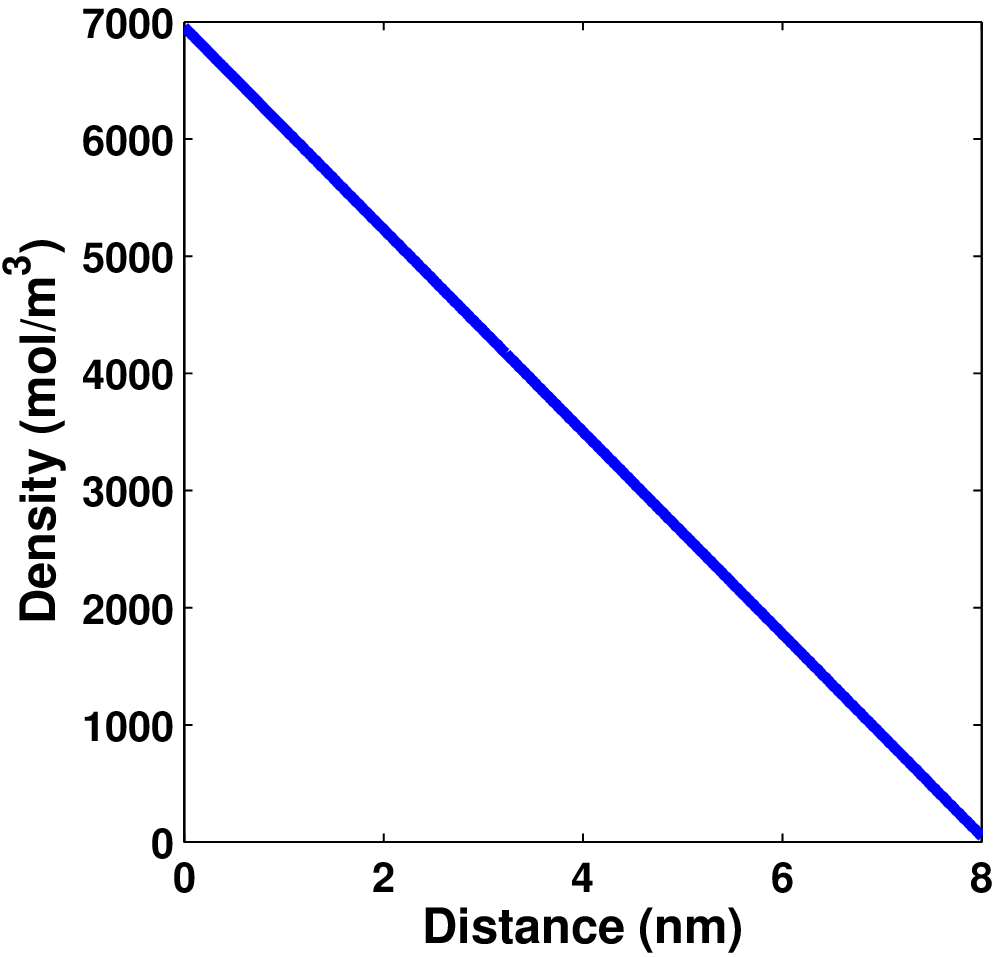}
        \caption{$s=0$.}
      \end{subfigure}
      \hfill
      \begin{subfigure}[t]{0.245\linewidth}
        \includegraphics[width=\linewidth]{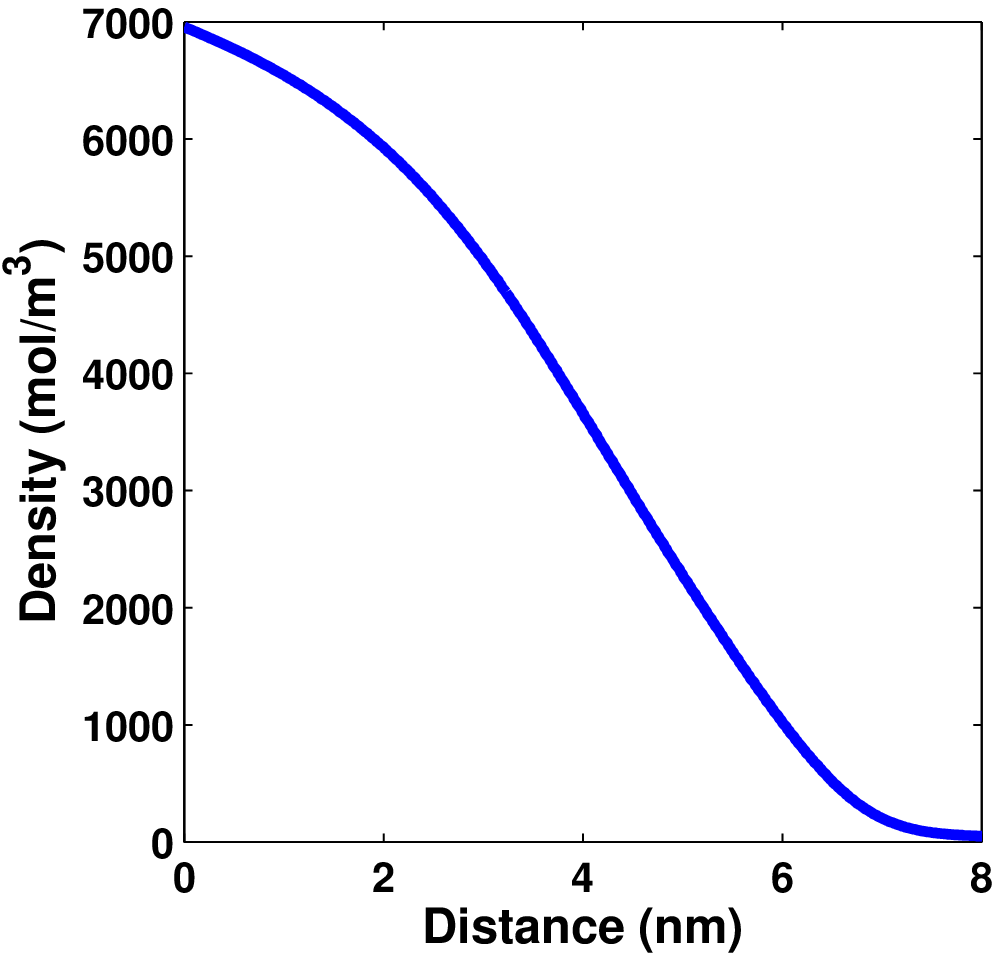}
        \caption{$s=0.2$.}
      \end{subfigure}
\hfill
      \begin{subfigure}[t]{0.245\linewidth}
        \includegraphics[width=\linewidth]{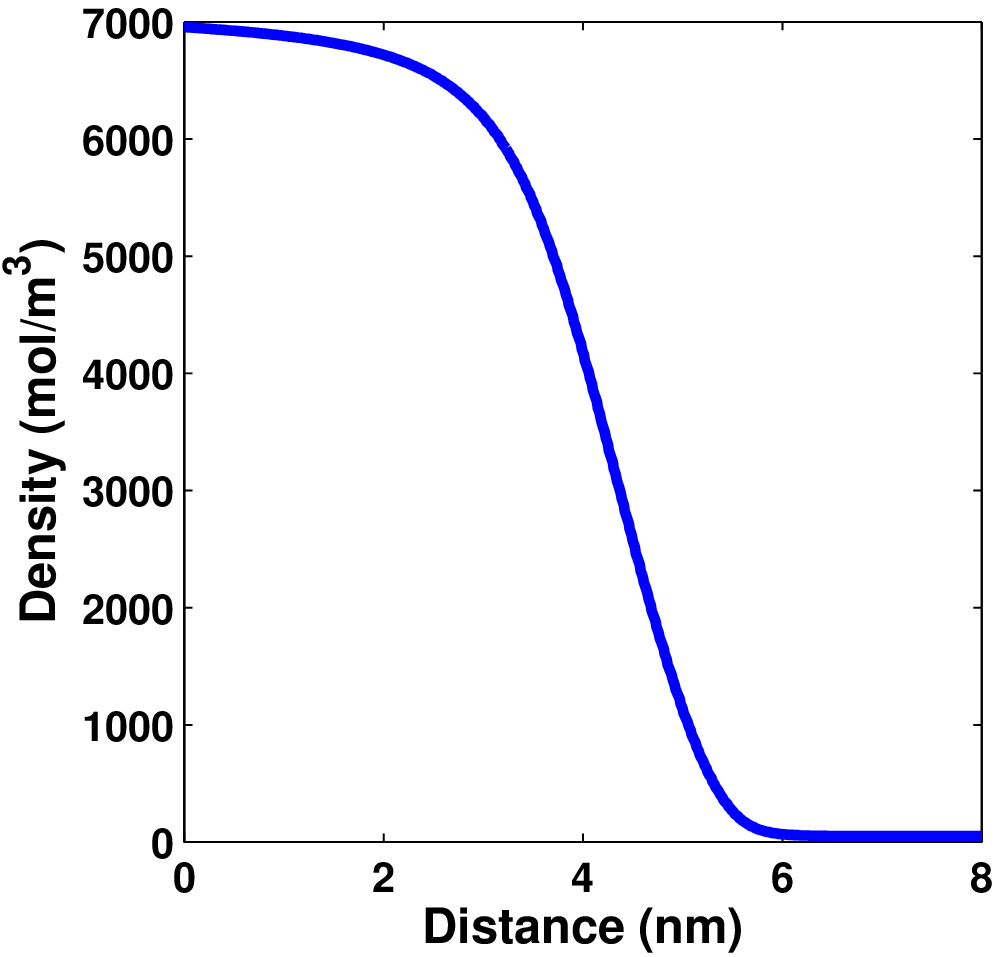}
        \caption{$s=0.6$.}
      \end{subfigure}
% FF: I removed two images since they are almost identical and there is no gain in showing these.
%       \begin{subfigure}[t]{0.475\linewidth}
%         \includegraphics[width=\linewidth]{hexane352Kt10}
%         \caption{$s=1$.}
%       \end{subfigure}
      \hfill
      \begin{subfigure}[t]{0.245\linewidth}
        \includegraphics[width=\linewidth]{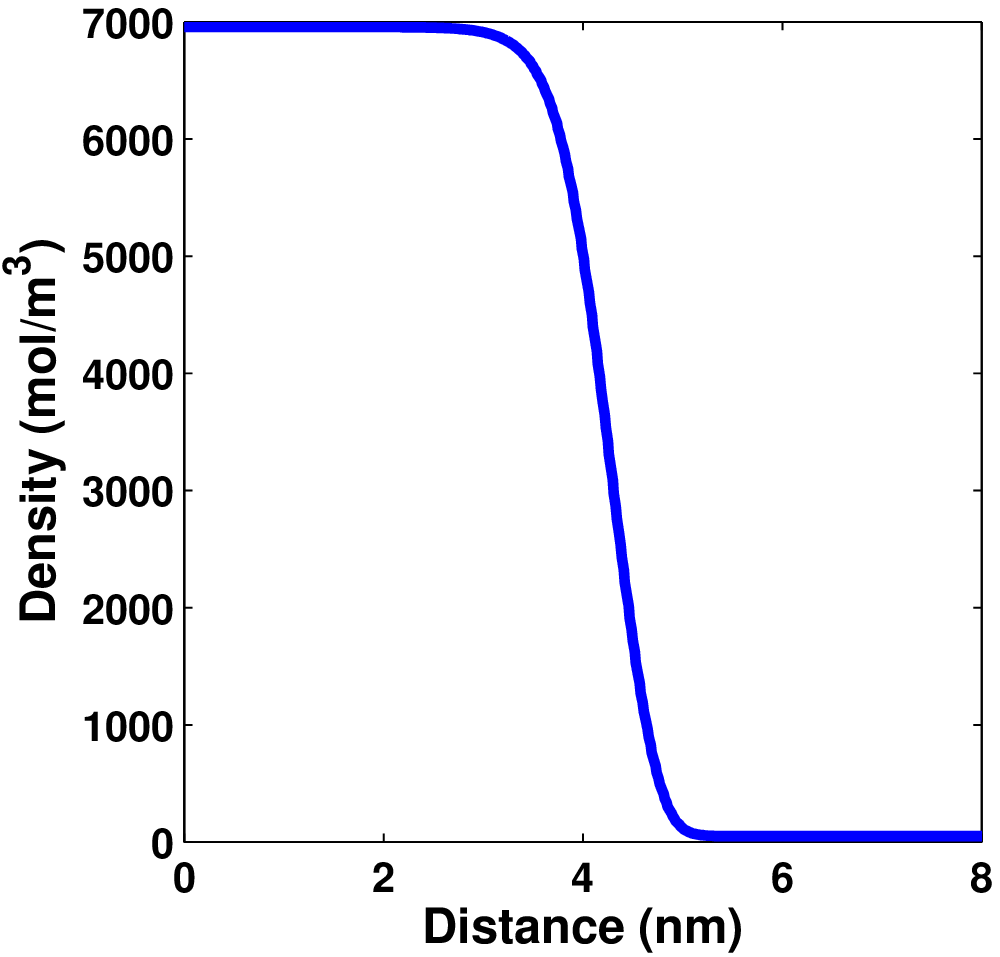}
        \caption{$s=4$.}
      \end{subfigure}
%       \begin{subfigure}[t]{0.475\linewidth}
%         \includegraphics[width=\linewidth]{hexane352Kt40}
%         \caption{$s=4$.}
%       \end{subfigure}
      \caption[The density profile of hexane (352.49~K) at different time step.]%
              {Density profile of hexane (352.49~K) at different time step ($\Delta s=0.1$). A~linear density distribution was used as initial condition at $s=0$ and the system reaches stable state at $s=4$.}
      \label{fig:dphexane352}
    \end{figure*}
    \par
  The SDGT algorithm uses the same boundary conditions as given in the original DGT, cf.~Eqn.~\eqref{eq:simplifiedEL1:bc} and~\eqref{eq:SDGT:BC}.  However, as opposed to the original DGT algorithm, the domain size~$D$ does not necessarily need to be close to the interface thickness~$L$ and any value that is larger than $L$ is suitable. This is validated through several numerical stability tests in the next section. Note that, although the first guess of~$D$ might not be large enough, the underestimation can be detected immediately from the converging behavior of the SDGT algorithm. Figure~\ref{fig:bcdiscussion}a shows the desired converging behavior when the prescribed~$D$ is larger than the interface thickness. After 10~time steps, densities close to the boundaries converge to bulk densities and the diffusive interface only occurs inside the domain. If the value of~$D$ is smaller than the interface thickness~$L$, diffusion will occur right next to the boundaries as shown in Figure~\ref{fig:bcdiscussion}b, which means the whole domain is inside the interface area. In this case, a~larger domain size is required to restart the iteration. It is usually safe to start from a~very large value of~$D$ for a~new system and switch to other values according to research purposes. We demonstrate below that the SDGT algorithm works quite well even on a~greatly overestimated domain size.
  
  %%%%%%%%%%%%%%%%%%%%%%%%%%%%%%%%%%%%%%%%%%%%%%%%%%%%%%%%%%%%%
  \section{Results and discussions}
  \label{sec:results}
  %%%%%%%%%%%%%%%%%%%%%%%%%%%%%%%%%%%%%%%%%%%%%%%%%%%%%%%%%%%%%
  The performance of the SDGT algorithm with \mbox{PC-SAFT} EoS was tested in several pure and mixed vapor-liquid equilibrium (VLE) systems from both physical and numerical aspects so as to validate the potential of applying this algorithm to perform predictive calculations.
\begin{figure}[ht!]
  \includegraphics[width=\linewidth]{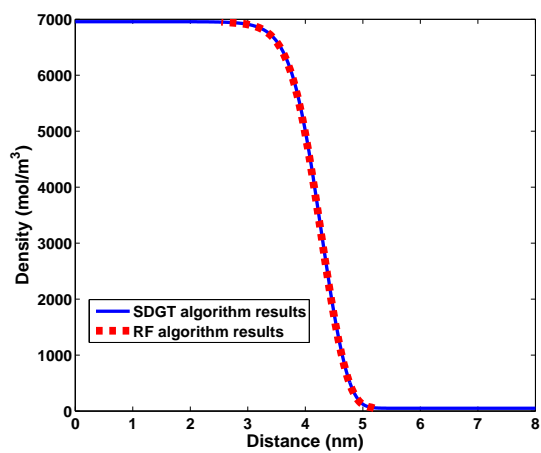}
  \caption[Calculated density profile of hexane at 352~K  by different algorithm. ]{Comparison of equilibrium density profile (hexane, 352.49~K): the SDGT algorithm (blue solid line) and RF algorithm (red dashed line).}
  \label{fig:compare_hexane}
\end{figure}

  \subsection{One component VLE system}
  Crude oil in reservoir is a multicomponent mixture consisting primarily of hydrocarbons \cite{pedersen2014phase}. Among different types of hydrocarbons, alkanes are of the most ones in crude oil. Therefore, a successful description of alkane interfacial properties will be of great interest to petroleum industry. Hexane~($C_6$), an alkane that consists six carbon atoms， was used in the first experiment. In Figure~\ref{fig:dphexane352}, the density profiles of hexane at various time steps are presented. The calculation started from a linear density distribution as initial condition on a 8~nm domain. With a time step of $\Delta s=0.1$, the system evolves quickly and steadily to reach an equilibrium state after 40 time steps when the density difference between two time steps meets the stopping criteria. 
    \begin{table}
      \caption{Numerical test results of the SDGT algorithm in one component system\textsuperscript{*}.}
      \label{tab:numericaltestpure}
        \begin{tabularx}{\linewidth}{@{}lCCC@{}}
          \toprule
          Density distribution  & $D~[\si{nm}]$ &   $\sigma~[\si{mN/m}]$  & AD (\%)
          \\\midrule
          Linear &  8 & 12.3557 &  0.127 \\ 
          Linear & 12 & 12.3562 & 0.131 \\ 
          Linear & 20 & 12.3580 & 0.146  \\ 
          Random & 10 & 12.3560 & 0.129\\ 
          \bottomrule
        \end{tabularx}
      \textsuperscript{*}~Experimental surface tension~$\sigma$ of hexane at 352.49~K: 12.34~mN/m~\cite{Grigoryev1992}. Interfacial thickness: 2.7~nm.
    \end{table}
  The equilibrium density profile (Figure~\ref{fig:dphexane352}d) is compared with the one given by the RF algorithm in Figure~\ref{fig:compare_hexane}. The two density profiles match in the interface region which proves that the SDGT algorithm can reproduce the equilibrium density profile. However, the computation of the RF algorithm is restricted in the interface region, while the SDGT algorithm provides additional information extending from the interface region to bulk phases. 
    \begin{figure}[ht!]
      \includegraphics[width=\linewidth]{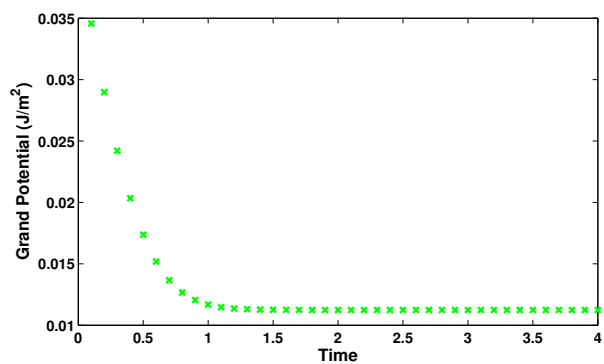}
      \caption[The dissipating process of grand potential during calculation.]%
              {The dissipating process of grand potential energy (hexane, 352.49~K) during calculation.}
      \label{fig:gphexane}
    \end{figure}
  During the calculation process, we also monitored the dissipation of the grand potential energy as shown in Figure~\ref{fig:gphexane}. It can be seen that the initial density distribution generates a very high energy environment which monotonically dissipates until the system reaches a stationary state.
  %More results of the equilibrium density profile from different systems are shown in Fig. (XXX).
  \begin{figure}[ht!]
    \begin{subfigure}[t]{0.9\linewidth}
      \includegraphics[width=\linewidth]{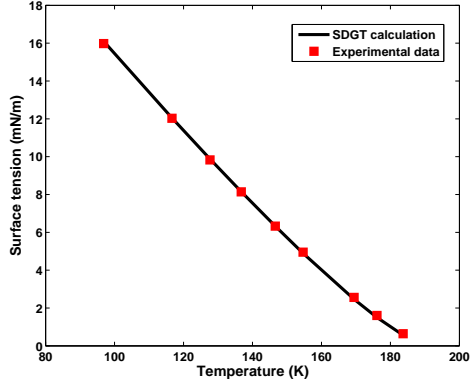}
      \caption{Methane.} 
    \end{subfigure}
    \hfill
    \begin{subfigure}[t]{0.9\linewidth}
      \includegraphics[width=\linewidth]{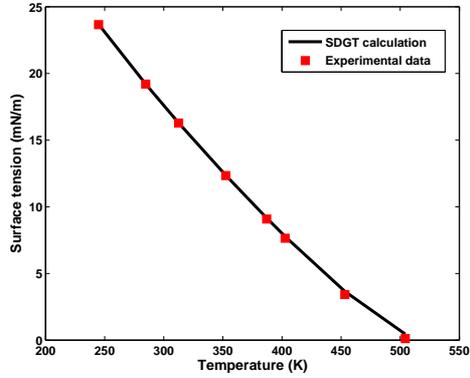}
      \caption{Hexane.} 
    \end{subfigure}
    \caption[Surface tension calculation results.]%
            {Comparison of the SDGT calculation results (blue solid line) with experimental data (red square dot) for surface tension: (a) Methane~\cite{Baidakov1982}, (b) Hexane~\cite{Grigoryev1992}.}
    \label{fig:pure_sigma}
  \end{figure}
  \begin{figure}[ht!]
    \centering
    \begin{subfigure}[t]{0.475\linewidth}
      \includegraphics[width=\linewidth]{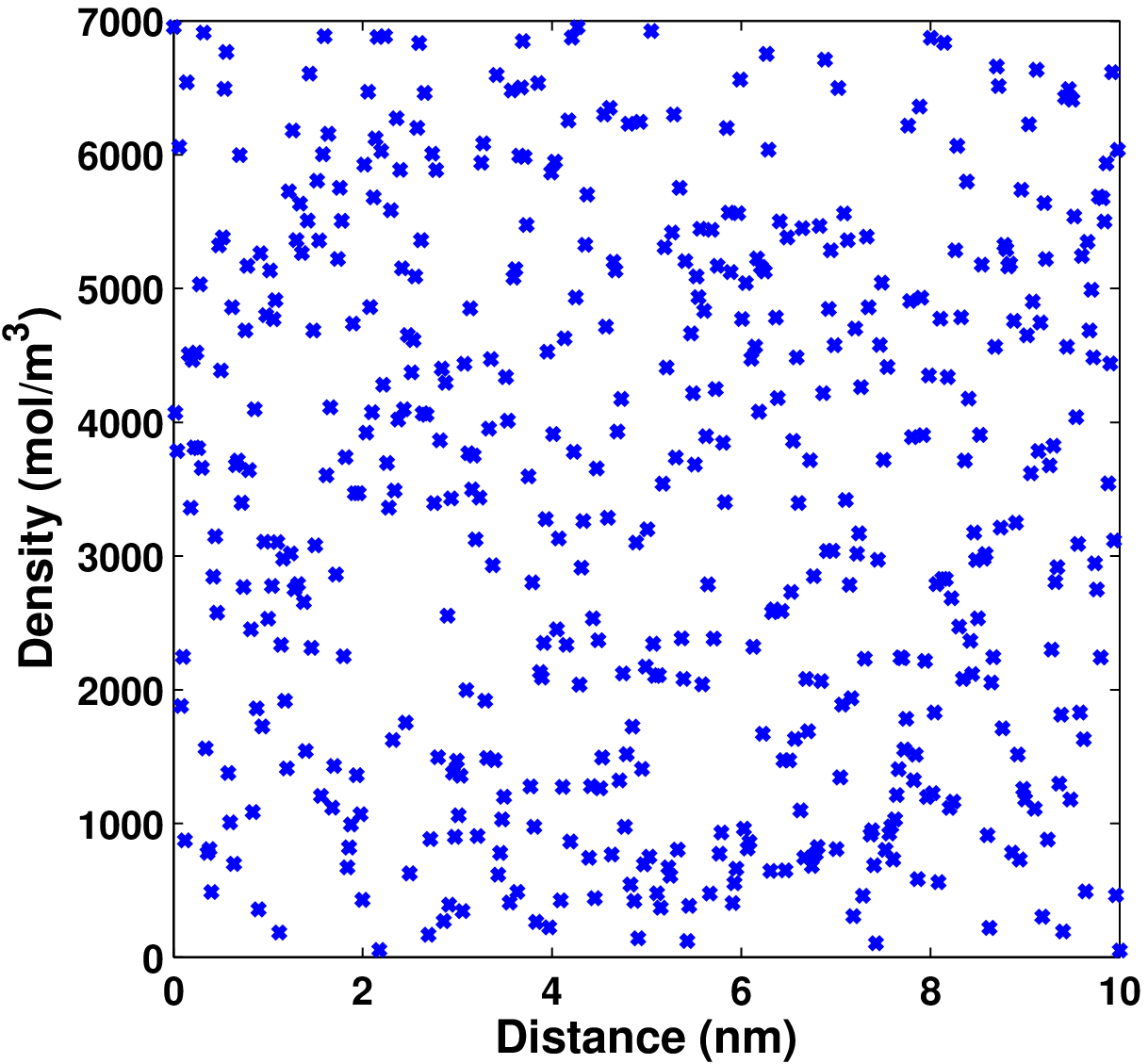}
      \caption{$s=0$.}
    \end{subfigure}
    \hfill
    \begin{subfigure}[t]{0.475\linewidth}
      \includegraphics[width=\linewidth]{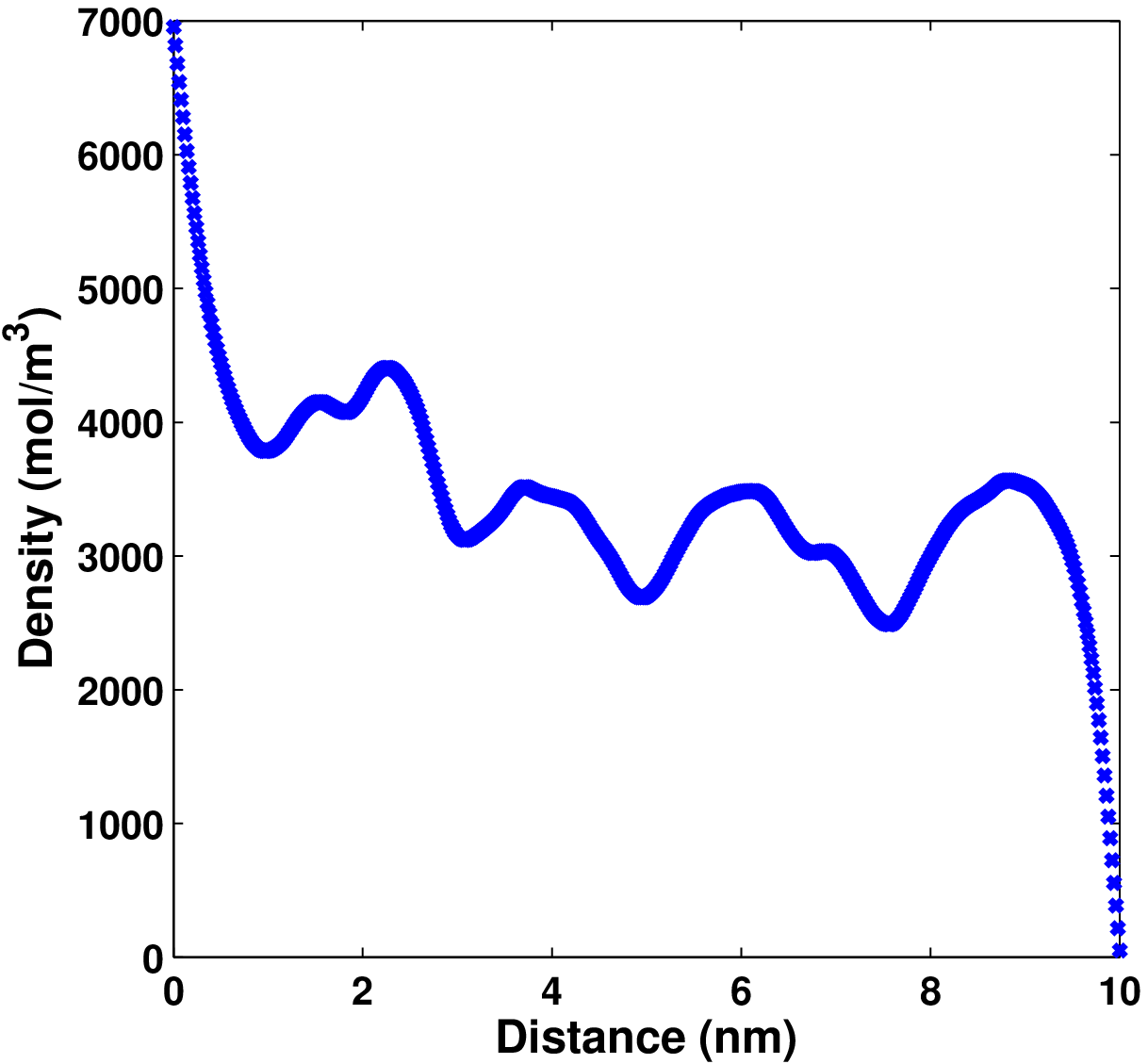}
      \caption{$s=1$.}
    \end{subfigure}
    \\
    \begin{subfigure}[t]{0.475\linewidth}
      \includegraphics[width=\linewidth]{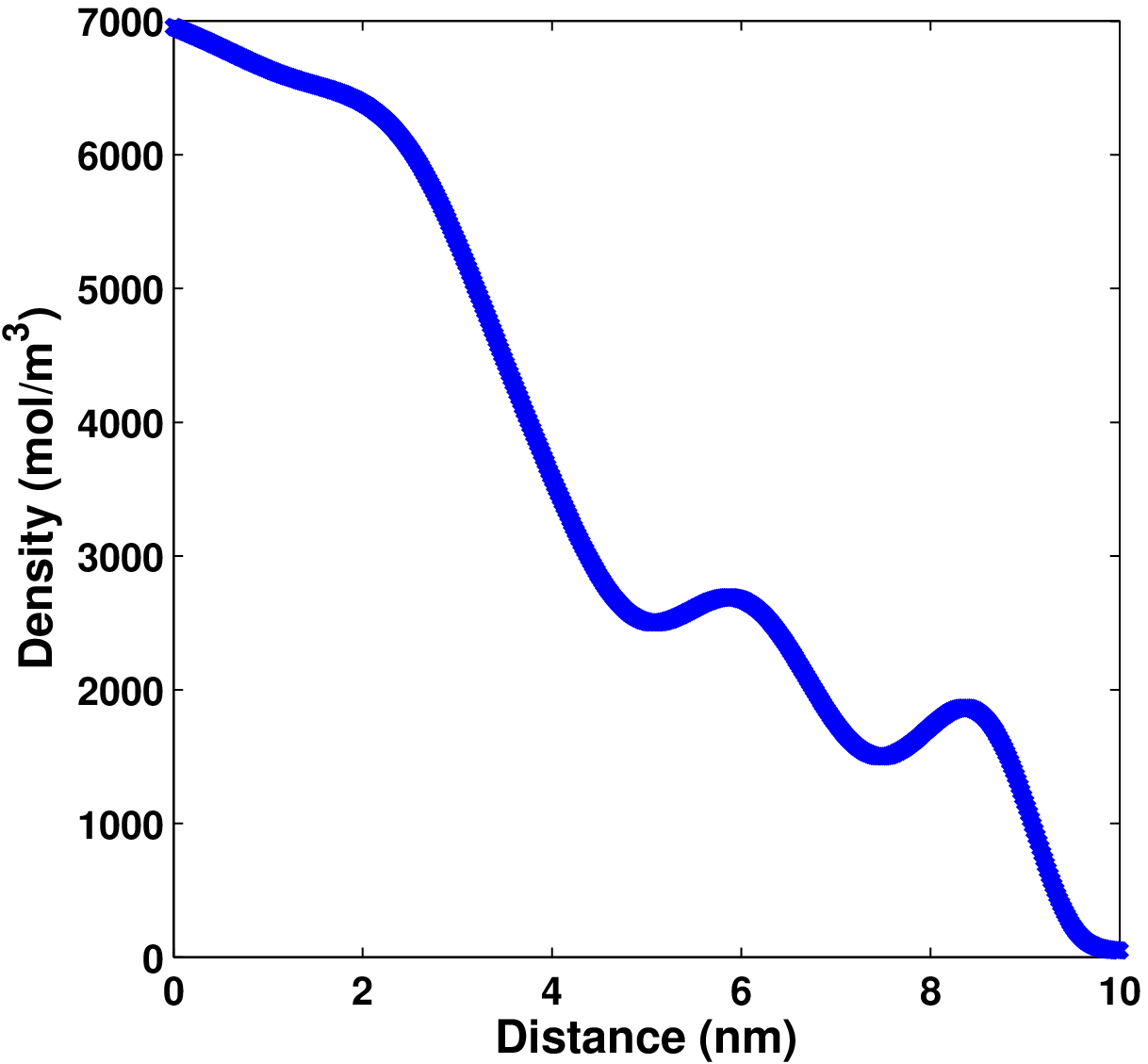}
      \caption{$s=5$.}
    \end{subfigure}
% FF: I removed two images since they are almost identical and there is no gain in showing these.
%     \begin{subfigure}[t]{0.475\linewidth}
%       \includegraphics[width=\linewidth]{hexane352Krandomt10}
%       \caption{$s=10$.}
%     \end{subfigure}
    \hfill
    \begin{subfigure}[t]{0.475\linewidth}
      \includegraphics[width=\linewidth]{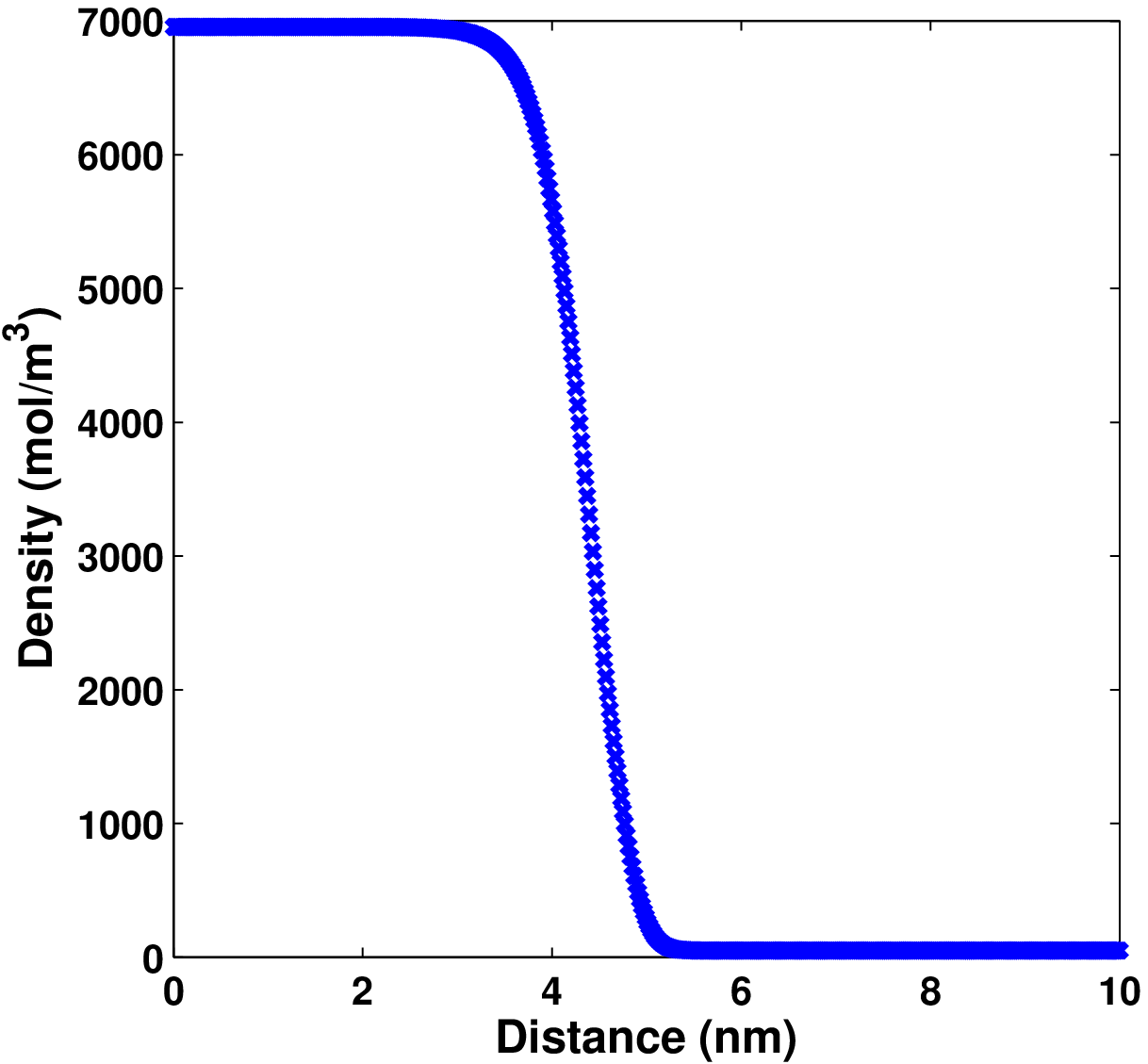}
      \caption{$s=20$.}
    \end{subfigure}
%     \begin{subfigure}[t]{0.475\linewidth}
%       \includegraphics[width=\linewidth]{hexane352Krandomt20}
%       \caption{$s=20$.}
%     \end{subfigure}
    \caption[Calculated density profile of hexane from a random initial guess.]%
            {Numerical stability test: use random density distribution as initial guess (hexane, 352.49~K).}
    \label{fig:randomhexane352}
  \end{figure}
  \par
  The surface tension of the system is calculated according to Eqn.~(\ref{eq:sigma1}) using the equilibrium density profiles. Calculation results are illustrated in Figure~\ref{fig:pure_sigma} as solid lines. The predicted surface tension decreases as a function of increasing temperature. This is because cohesive forces between molecules decreases with an increase of system thermal energy. The surface tension trend and values are further verified by comparing with experimental data (rectangular dots) \cite{Baidakov1982,Grigoryev1992}. Excellent agreements are observed throughout a wide temperature range for both systems. 
  \begin{figure*}[ht!]
    \centering
    \begin{subfigure}[t]{0.245\linewidth}
      \includegraphics[width=\linewidth]{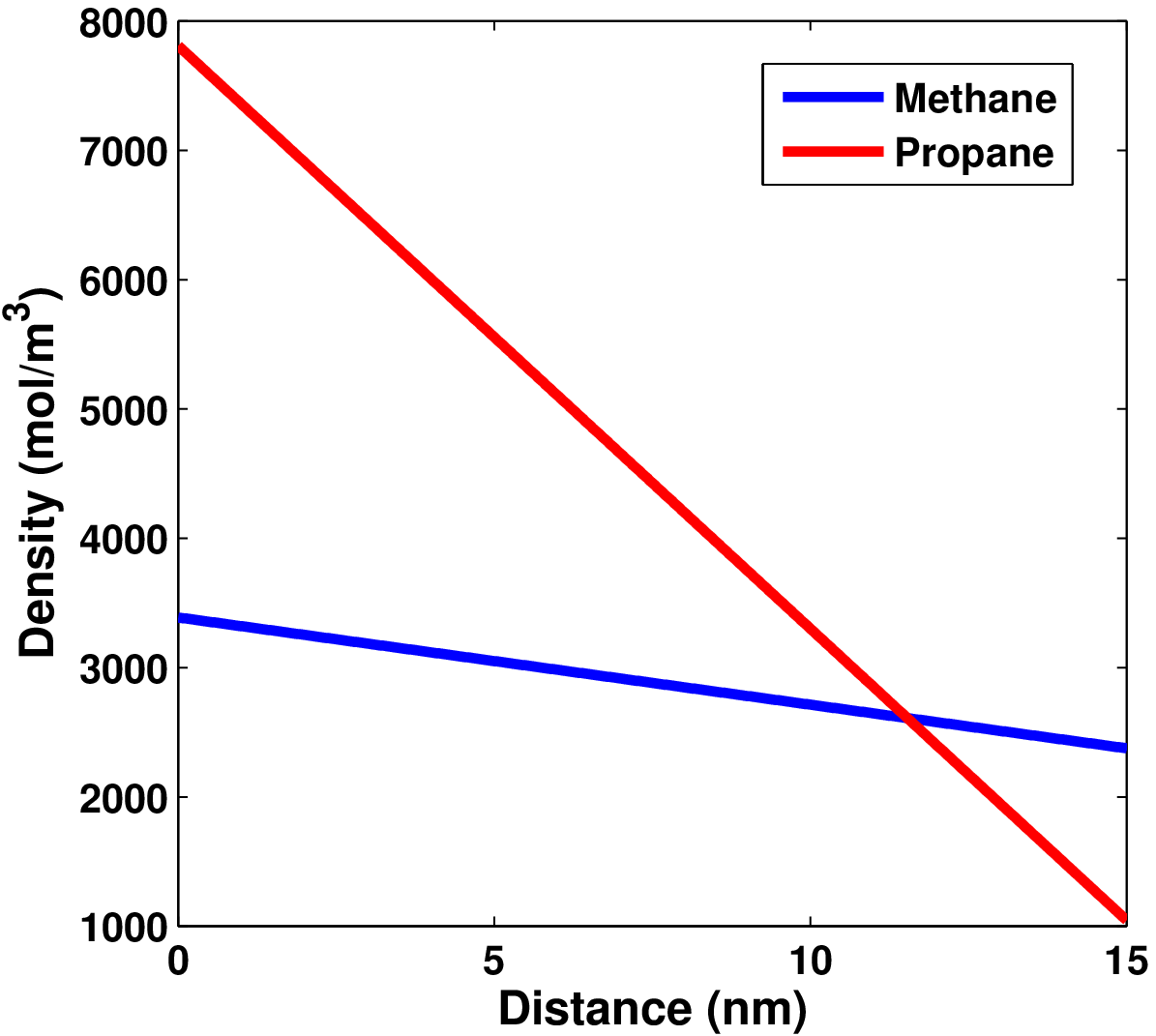}
      \caption{$s=0$.}
    \end{subfigure}
    \hfill
    \begin{subfigure}[t]{0.245\linewidth}
      \includegraphics[width=\linewidth]{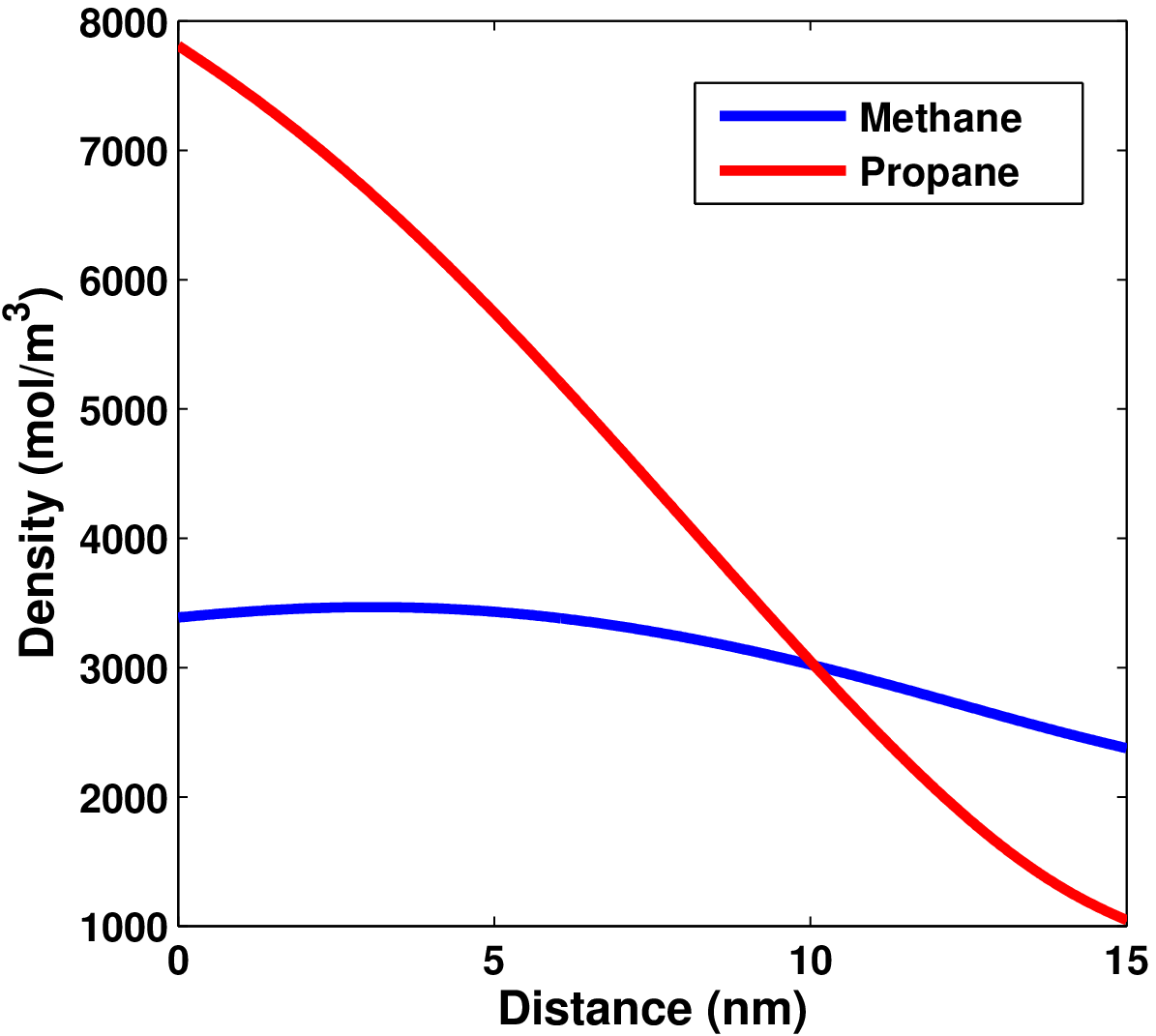}
      \caption{$s=1$.}
    \end{subfigure}
    \hfill
    \begin{subfigure}[t]{0.245\linewidth}
      \includegraphics[width=\linewidth]{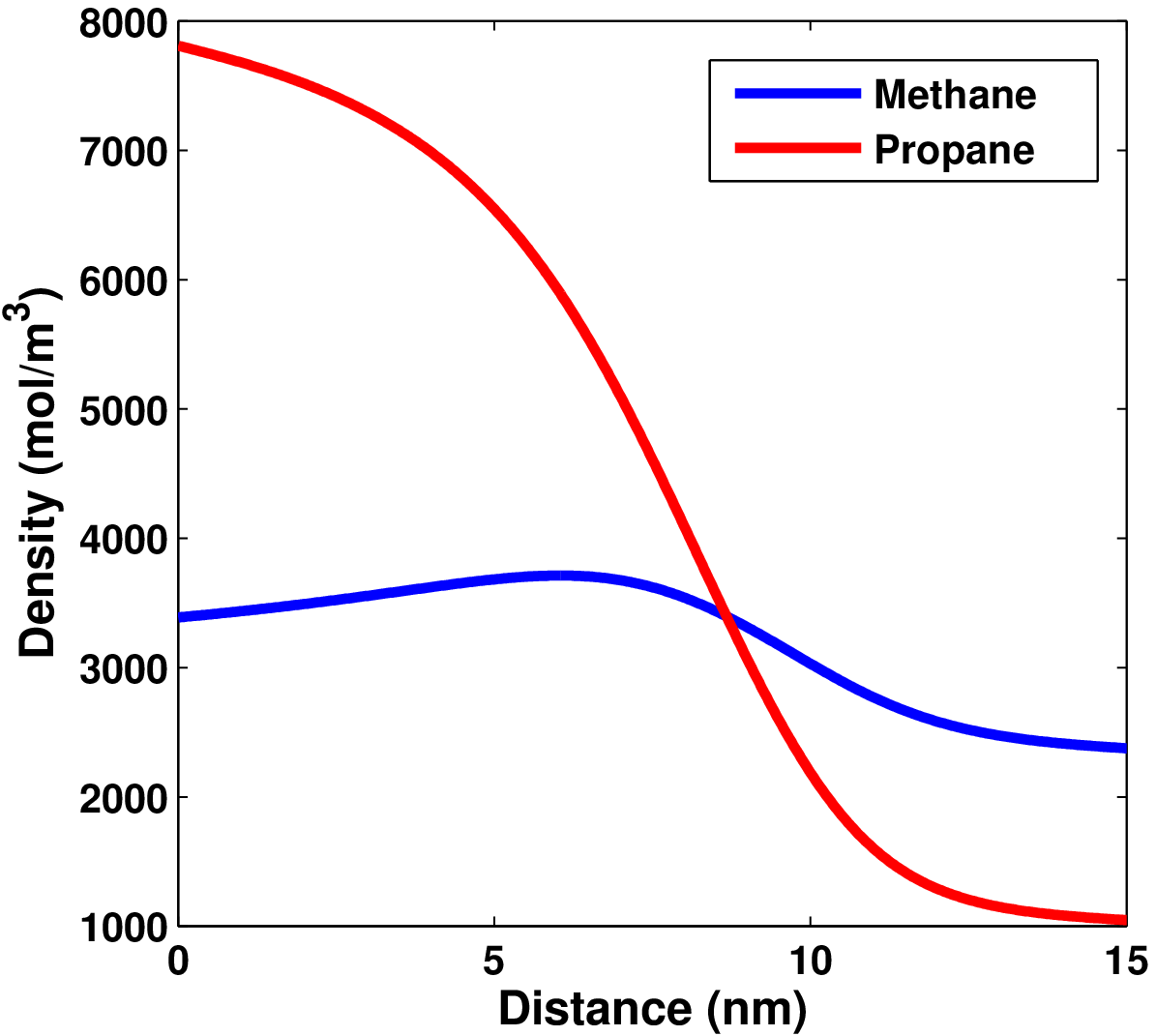}
      \caption{$s=5$.}
    \end{subfigure} 
    \hfill
%     \begin{subfigure}[t]{0.475\linewidth}
%       \includegraphics[width=\linewidth]{methane_propane_303K_60bar_t10}
%       \caption{$s=10$.}
%     \end{subfigure}
    \hfill
    \begin{subfigure}[t]{0.245\linewidth}
      \includegraphics[width=\linewidth]{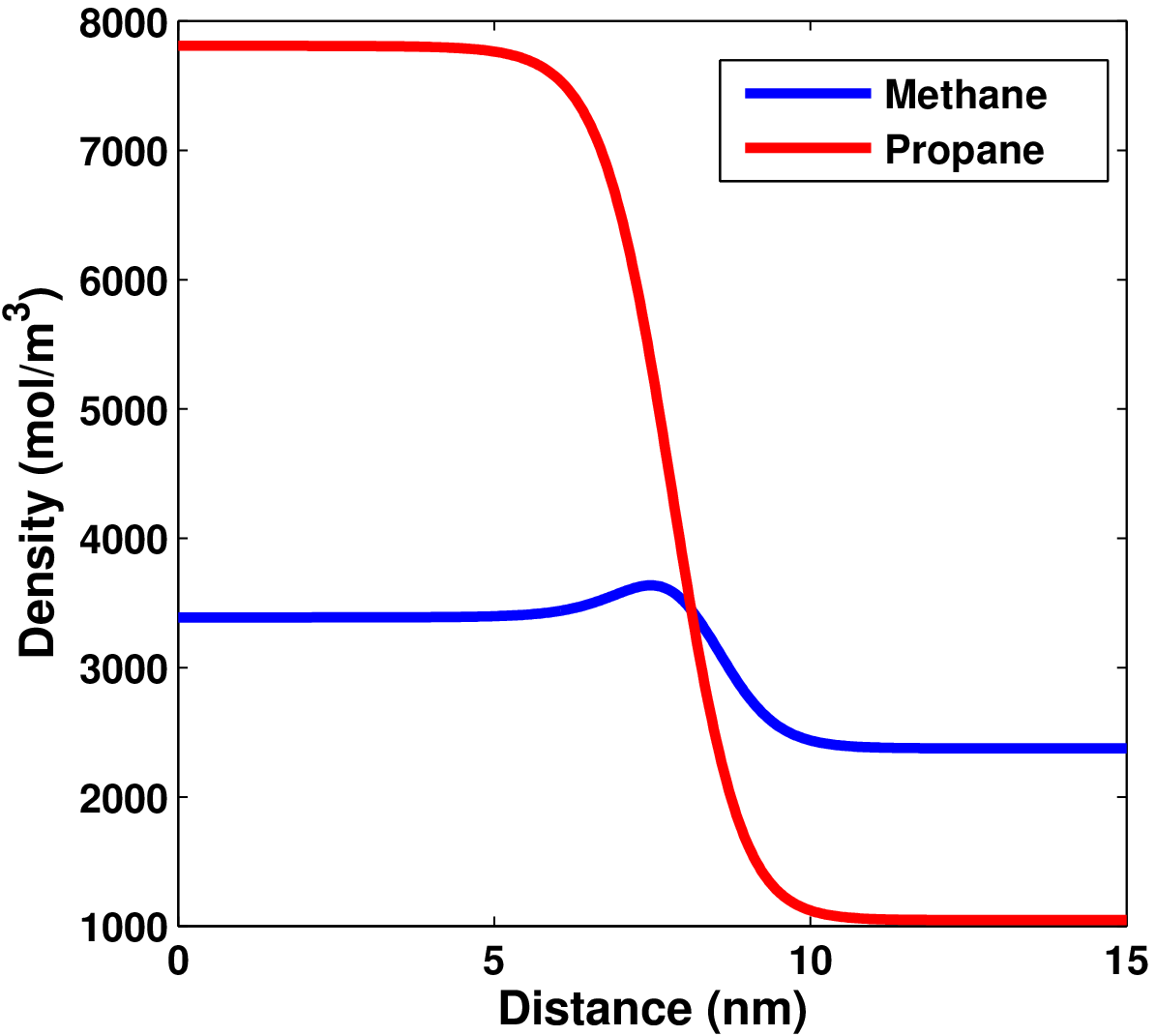}
      \caption{$s=30$.}
    \end{subfigure}
%     \hfill
%     \begin{subfigure}[t]{0.475\linewidth}
%       \includegraphics[width=\linewidth]{methane_propane_303K_60bar_t30}
%       \caption{$s=30$.}
%     \end{subfigure}
    \caption[Density profile of methane--propane mixture at 303.15~K and 60~bar.]%
            {Density profile of methane--propane mixture (303.15~K, 60~bar) at different time ($\Delta s=1$). A linear density distributions for both components were used as initial condition at $s=0$ and the system reaches equilibrium at $s=30$.}
    \label{fig:hexanepropane303K60bar}  
  \end{figure*}
  \par
  We also conducted numerical stability and robustness tests on the SDGT algorithm. In the first test case, a random density distribution is used as initial condition, in which local densities are produced by a random number generator bounded by bulk densities, as shown in Figure~\ref{fig:randomhexane352}a. This is an extreme situation in which different phases in the system are evenly mixed. The converging process from Figure~\ref{fig:randomhexane352}a to Figure~\ref{fig:randomhexane352}d demonstrates the robustness of this algorithm dealing with such a suboptimal condition. Without being subjected to a mass conservation, the system first converges to an oscillatory but smooth line, and then reaches its equilibrium state after 20 time steps. The surface tension result of this test is reported and compared with experimental data in Table \ref{tab:numericaltestpure} with a high accuracy. This test  verifies our statement in Section \ref{sec:initialguess} that the SDGT algorithm is numerically flexible with initial conditions which are usually difficult to estimate. In practical calculations, a linear density distribution will function as a simple and adequate initial condition.
    \begin{table}
      \caption{Numerical test results of the SDGT algorithm in binary mixture system\textsuperscript{*}.}
      \label{tab:numericaltestmixture}
        \begin{tabularx}{\linewidth}{@{}lCCC@{}}
          \toprule
          Density distribution & $D~[\si{nm}]$ & $\sigma~[\si{mN/m}]$  & AD~(\%)\\
          \midrule
           Linear & 10 &   2.1789 &  0.127 \\ 
           Linear & 15 &  2.1796  & 0.131 \\ 
           Linear & 25 &  2.1796  & 0.146  \\ 
           Random & 10 &  2.1799  & 0.129\\ 
          \bottomrule
        \end{tabularx}
        \textsuperscript{*}~Experimental surface tension~$\sigma$ of methane--propane at 303.15~K, 60~bar: 2.14~mN/m~\cite{weinaug1943surface}. Interfacial thickness: 8.09~nm.
    \end{table}
    
  Another test is solving for interfacial properties of the same system on different domain sizes. From Figure~\ref{fig:compare_hexane}, it can be concluded that the interface thickness of hexane at 352.49 K is 2.7 nm. Table \ref{tab:numericaltestpure} reports the surface tension calculation results of hexane on a domain size of 8 nm, 12 nm and 20 nm at 352.49 K.  By using the SDGT algorithm, even given a much wider domain size (20 nm for example), the system still converges steadily to generate an accurate surface tension result, which further demonstrates the robustness of the SDGT algorithm on different boundary conditions as stated in Section \ref{sec:spacedis}.

  \subsection{Multi-component VLE system}
  In our work, the SDGT algorithm was for the first time applied to a multicomponent system. Compared with the RF algorithm, no reference substance is needed to start the calculation. Crossing influence parameters can be calculated by a mixing rule of pure component influence parameters:
  \begin{equation}\label{eq:mixingrule:vij}
    v_{ij} = (1-\beta_{ij})\,\sqrt{v_i\,v_j}\,,
  \end{equation}
  where $\beta_{ij}$ is the binary interaction parameter of influence parameter.
  
  A methane--propane mixed system was used to test the performance of the SDGT algorithm for mixtures. Figure~\ref{fig:hexanepropane303K60bar} shows the density profile solving process of the mixture at 303.15~K, 60~bar. Starting from a linear density distribution (Figure~\ref{fig:hexanepropane303K60bar}a), the density profiles evolve gradually and reach the equilibrium distribution after 30 time steps as shown in Figure~\ref{fig:hexanepropane303K60bar}d. An adsorption of methane on the propane rich liquid surface is observed in the equilibrium density profile. This is to keep the minimum system free energy which can be better illustrated on a free energy contour map.
  \begin{figure}[ht!]
  	\centering
  	\includegraphics[width=\linewidth]{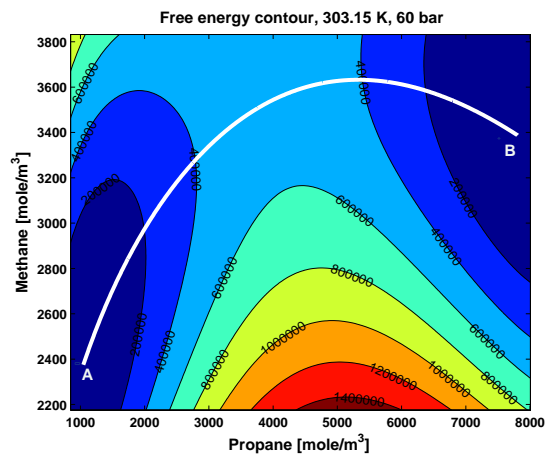}
  	\caption{The free energy contour of methane--propane mixture at 303.15 K, 60 bar. The while line is the equilibrium density profile from the SDGT calculations.}
  	\label{fig:free_energy_contour}
  \end{figure}
  \par
  Figure~\ref{fig:free_energy_contour} shows a contour map of the methane--propane system free energy as a function of their densities. This free energy surface is generated by subtracting the tangent plane from the Helmholtz free energy surface~\cite{rowlinson2013molecular}. The white path line is plotted on the contour map using the equilibrium density profile from SDGT calculations. It can be observed that in order to minimize the free energy along the path from the vapor rich phase (point A) to the liquid rich phase (point B), the density of methane must increase toward the saddle point of the energy surface so as to avoid the high energy hills, resulting in the surface accumulation of methane. A same description was presented by Sergio Cisneros et al.~\cite{Sergio19th}. Besides, this methane surface adsorption is also validated by molecular simulations~\cite{herdes2015coarse}.
  \begin{figure}[ht!]
  	\centering
  	\includegraphics[width=\linewidth]{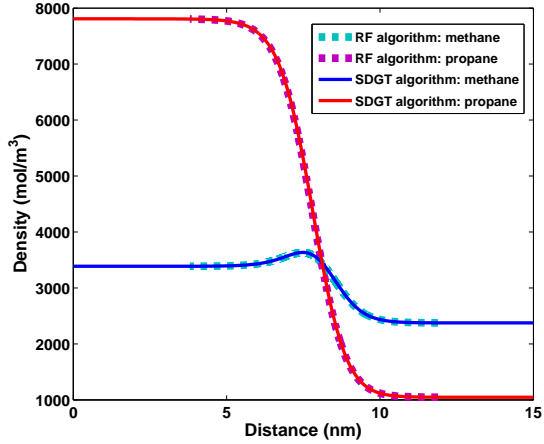}
  	\caption[Compare density profile calculation of different approach in mixture system.]{Comparison of equilibrium density profile (methane--propane system, 303.15 K, 60 bar): the SDGT algorithm (solid line) and the RF algorithm (dashed line).}
  	\label{fig:compare_dp_VLE}
  \end{figure}
  \par  
  The equilibrium density profiles from the SDGT algorithm (cf.~Figure~\ref{fig:hexanepropane303K60bar}d) are compared with the results from the RF algorithm. A~remarkable agreement between the two density profiles in the interface region is seen in Figure~\ref{fig:compare_dp_VLE} where solid lines represent results from the SDGT algorithm and dashed lines represent results from the RF algorithm. Again, the calculation of the RF algorithm stops immediately when it reaches the bulk phases while the SDGT algorithm presents the density distribution in the whole domain, including both the interface region and bulk region.
  \begin{figure}[ht!]
  	\centering
  	\includegraphics[width=\linewidth]{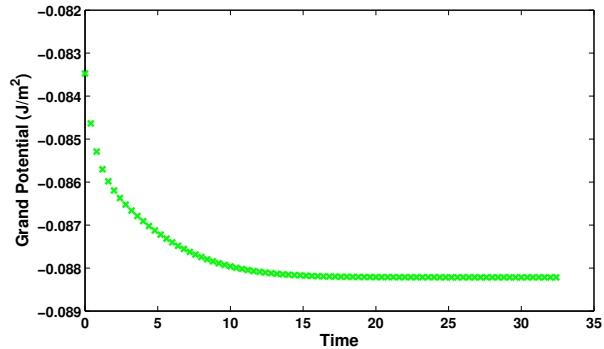}
  	\caption[The dissipating process of grand potential in methane--propane system.]%
  	{The dissipating process of grand potential energy (methane--propane, 303.15~K, 60~bar) during calculation.}
  	\label{fig:gp_methane_propane}
  \end{figure}
  \par 
  The grand potential energy of the mixture was recorded during the evolutionary convergence process and results are displayed in Figure~\ref{fig:gp_methane_propane}. The energy maintains a high value at the initial condition and decreases monotonically afterwards. It reaches a minimum where the system achieves equilibrium.

  \begin{figure}[ht!]
  	\centering
  	\begin{subfigure}[t]{0.9\linewidth}
  		\includegraphics[width=\linewidth]{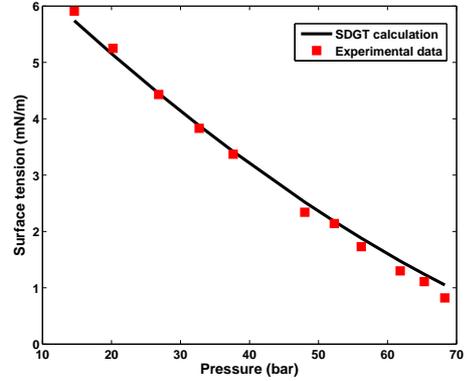}
  		\caption{Methane--propane, $T= 303.15~\mathrm{K}$.} 
  	\end{subfigure}
  	\hfill
  	\begin{subfigure}[t]{0.9\linewidth}
  		\includegraphics[width=\linewidth]{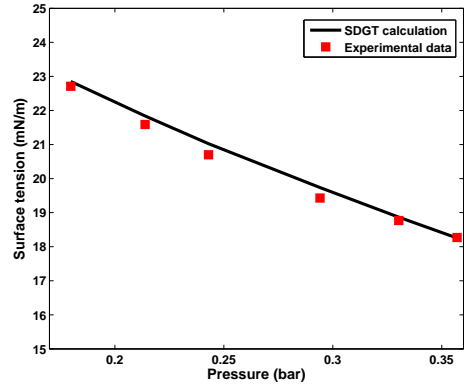}
  		\caption{Pentane--toluene, $T= 288.15~\mathrm{K}$.} 
  	\end{subfigure}
  	\caption[Surface tension calculation results.]%
  	{Comparison of the SDGT calculation results (blue solid line) with experimental data (red square dot) for surface tension: (a) Methane--propane~\cite{weinaug1943surface}, (b) Pentane--toluene\cite{Mahl1972}.}
  	\label{fig:mixture_sigma}
  \end{figure}
  \par
  Based on the equilibrium density profile, surface tension was calculated and plotted together with experimental data for methane--propane mixture (Figure~\ref{fig:mixture_sigma}a) \cite{weinaug1943surface} and pentane--toluene mixture (Figure~\ref{fig:mixture_sigma}b) \cite{Mahl1972}. Results are promising and it reveals that the SDGT algorithm operates in mixture systems with satisfactory accuracy as well.
  \par
\begin{figure}[ht!]
    \centering
    \begin{subfigure}[t]{0.475\linewidth}
      \includegraphics[width=\linewidth]{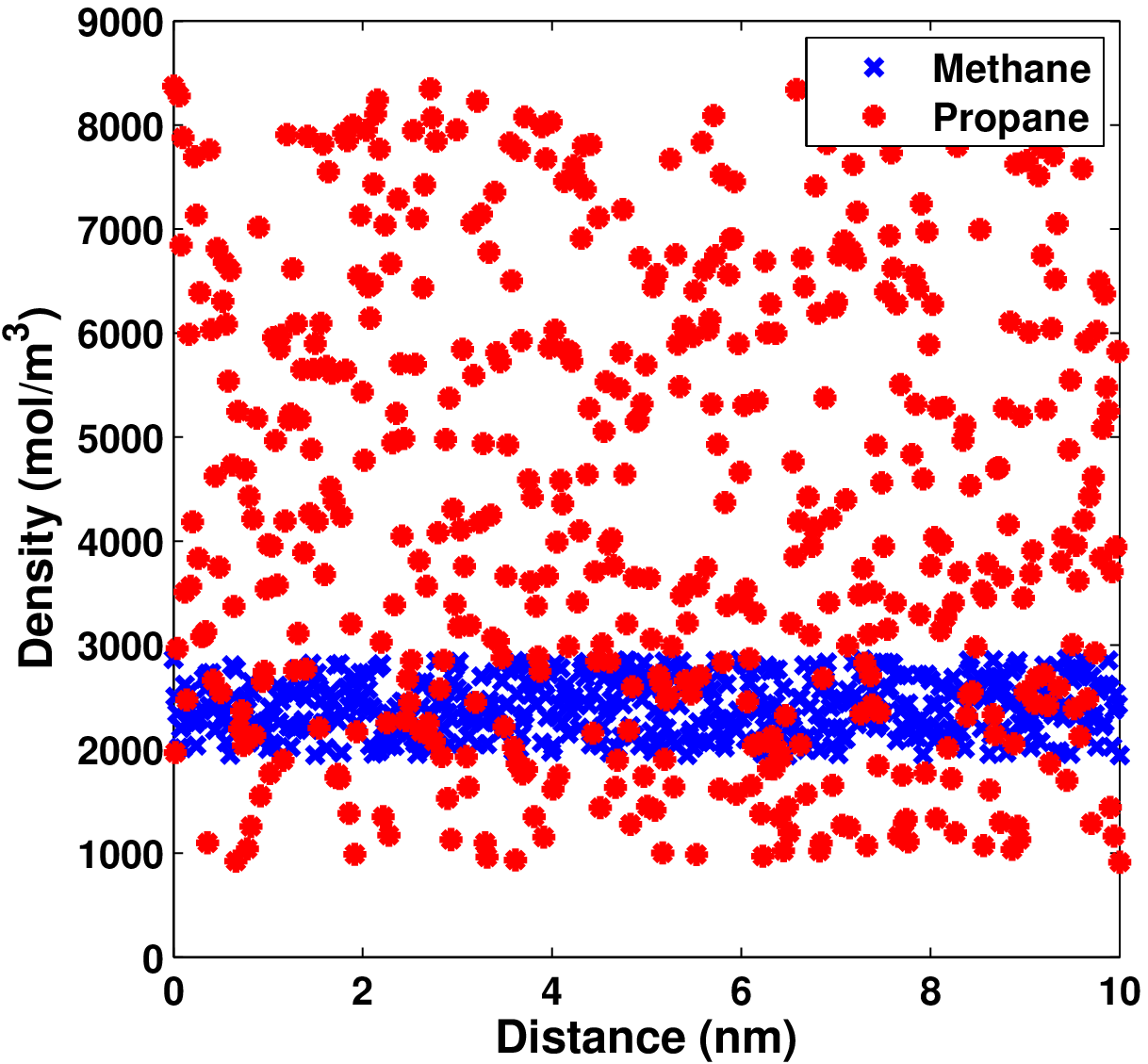}
      \caption{$s=0$.}
    \end{subfigure}
    \hfill
    \begin{subfigure}[t]{0.475\linewidth}
      \includegraphics[width=\linewidth]{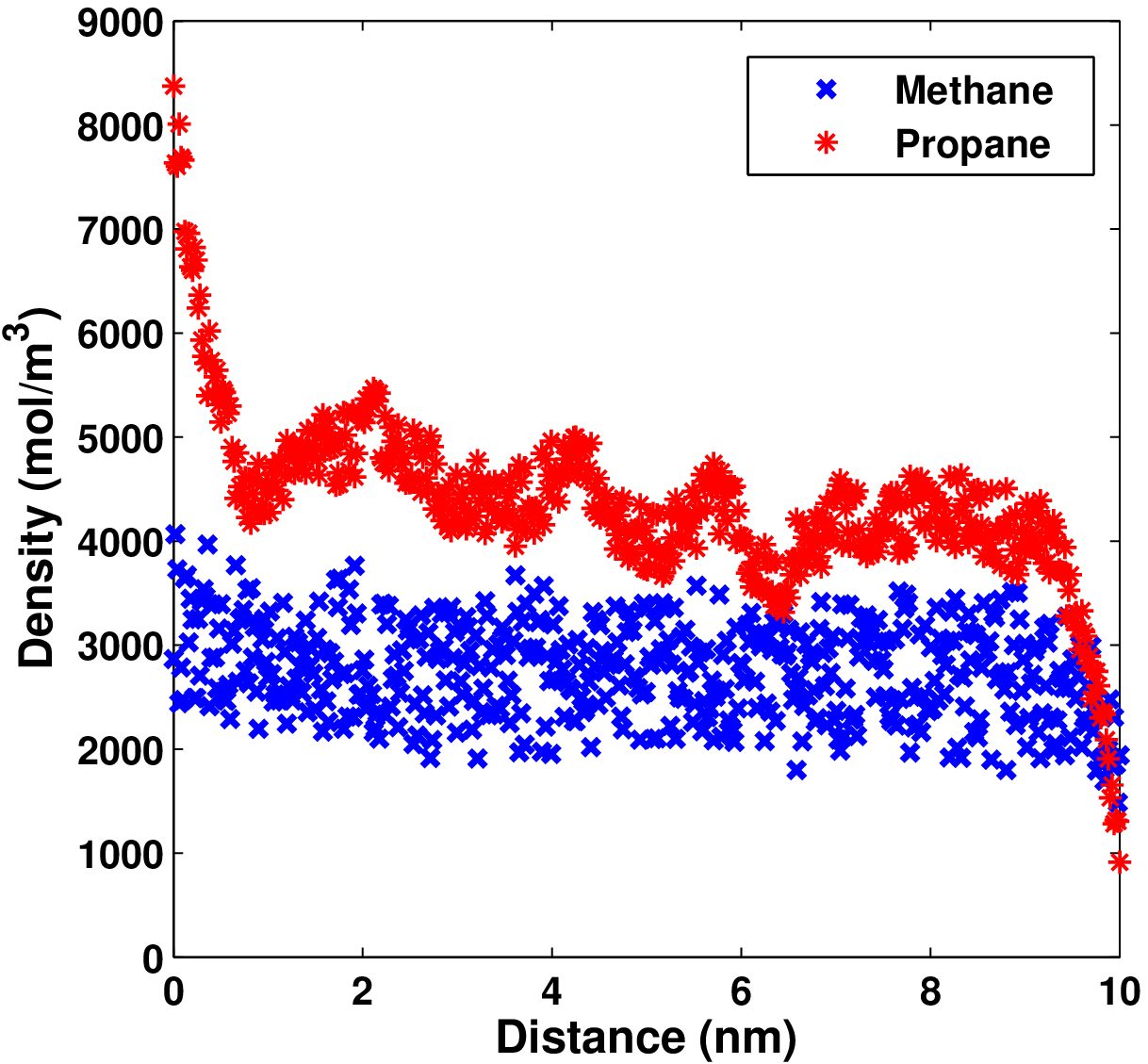}
      \caption{$s=1$.}
    \end{subfigure}
    \\
    \begin{subfigure}[t]{0.475\linewidth}
      \includegraphics[width=\linewidth]{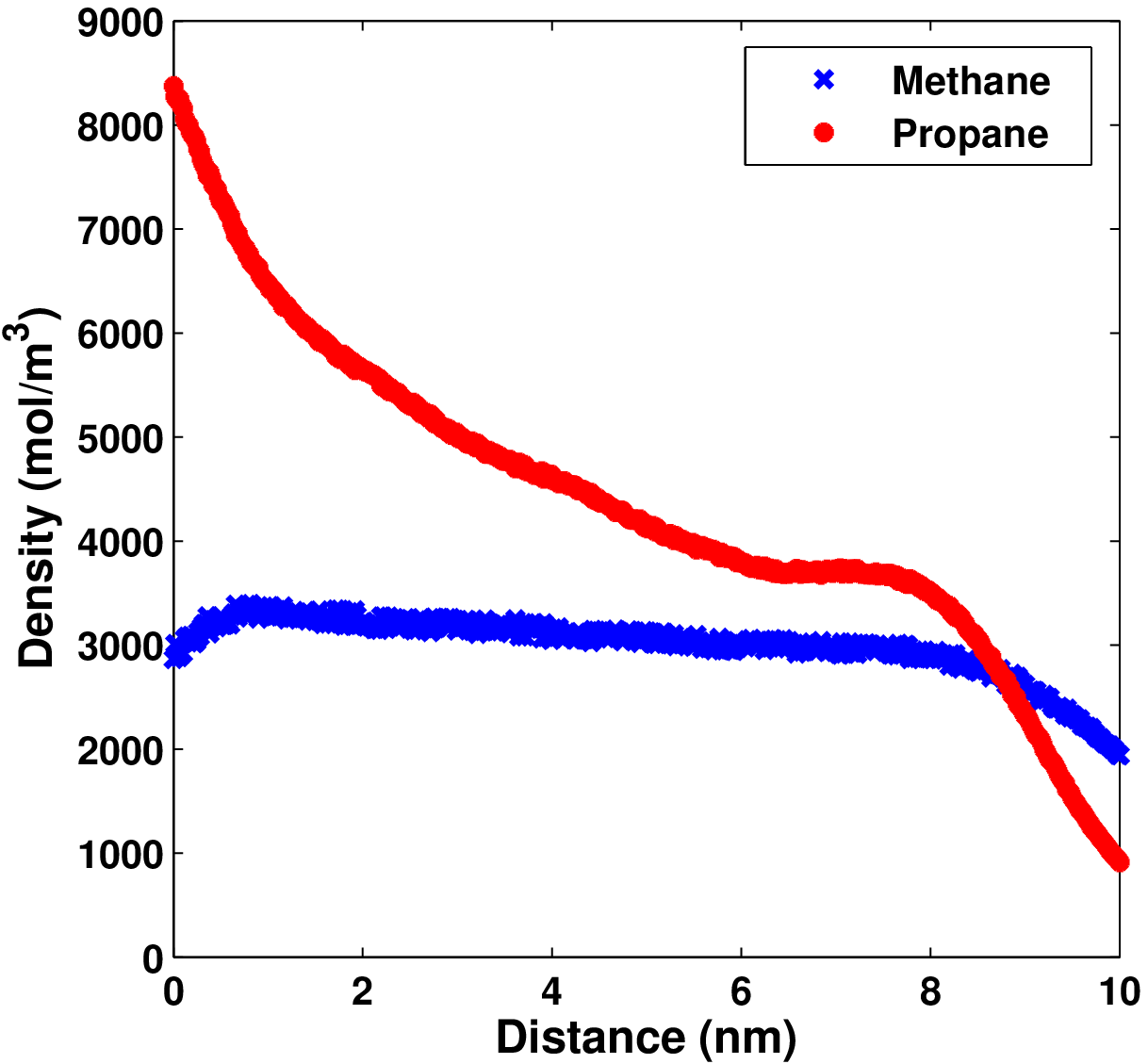}
      \caption{$s=5$.}
    \end{subfigure}
    \hfill
    \begin{subfigure}[t]{0.475\linewidth}
      \includegraphics[width=\linewidth]{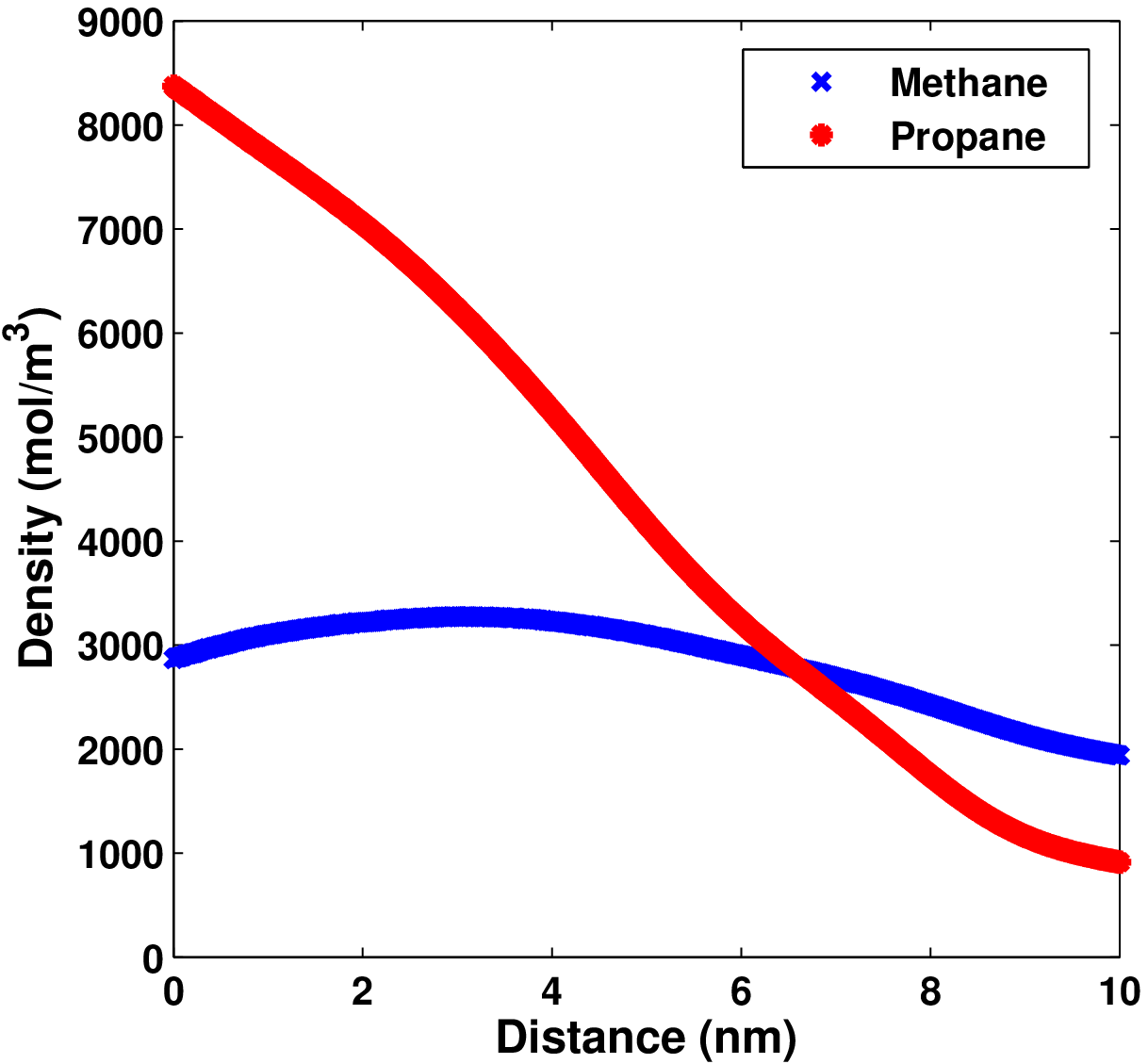}
      \caption{$s=10$.}
    \end{subfigure}
    \\
    \begin{subfigure}[t]{0.475\linewidth}
      \includegraphics[width=\linewidth]{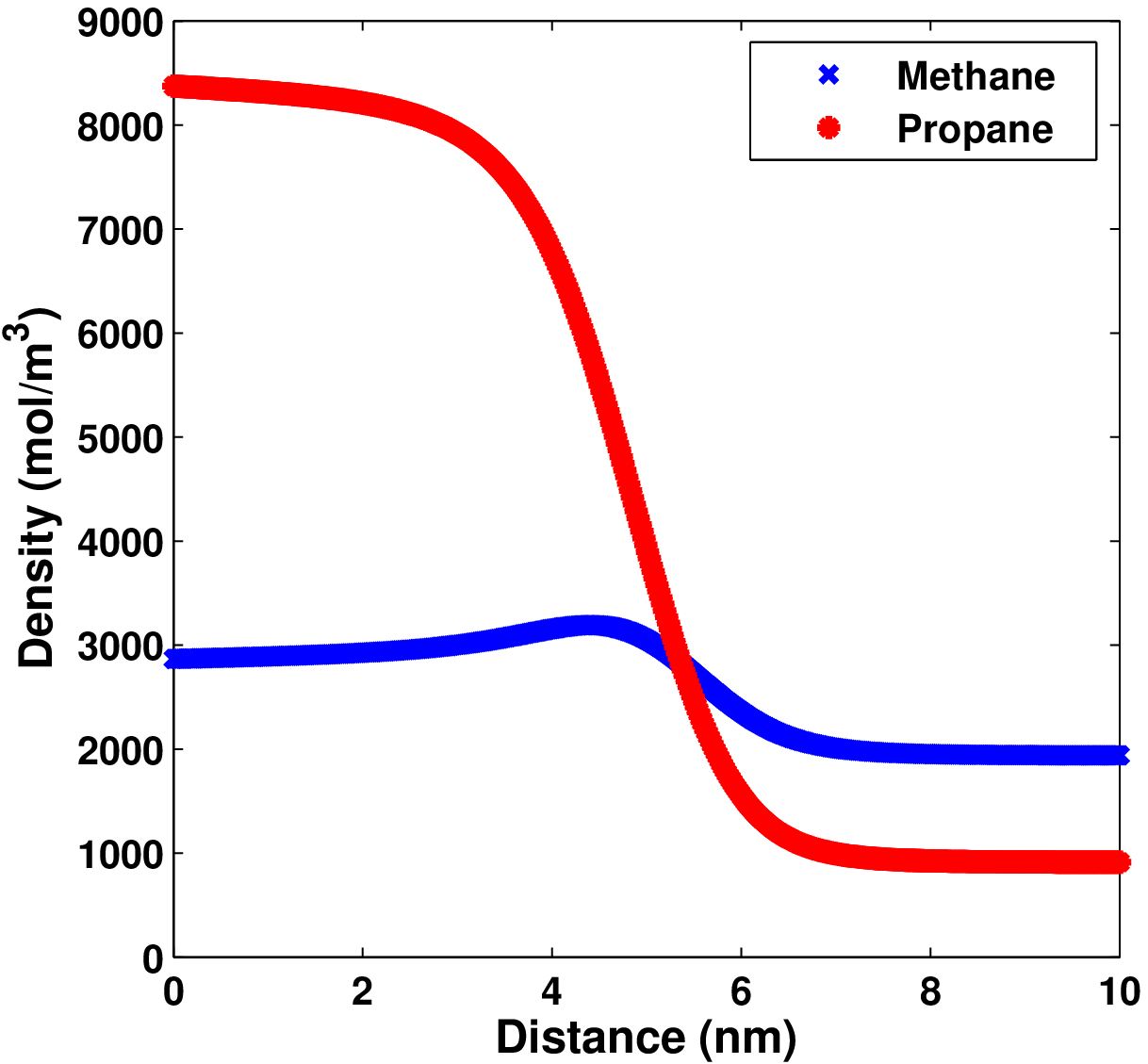}
      \caption{$s=20$.}
    \end{subfigure}
    \hfill
    \begin{subfigure}[t]{0.475\linewidth}
      \includegraphics[width=\linewidth]{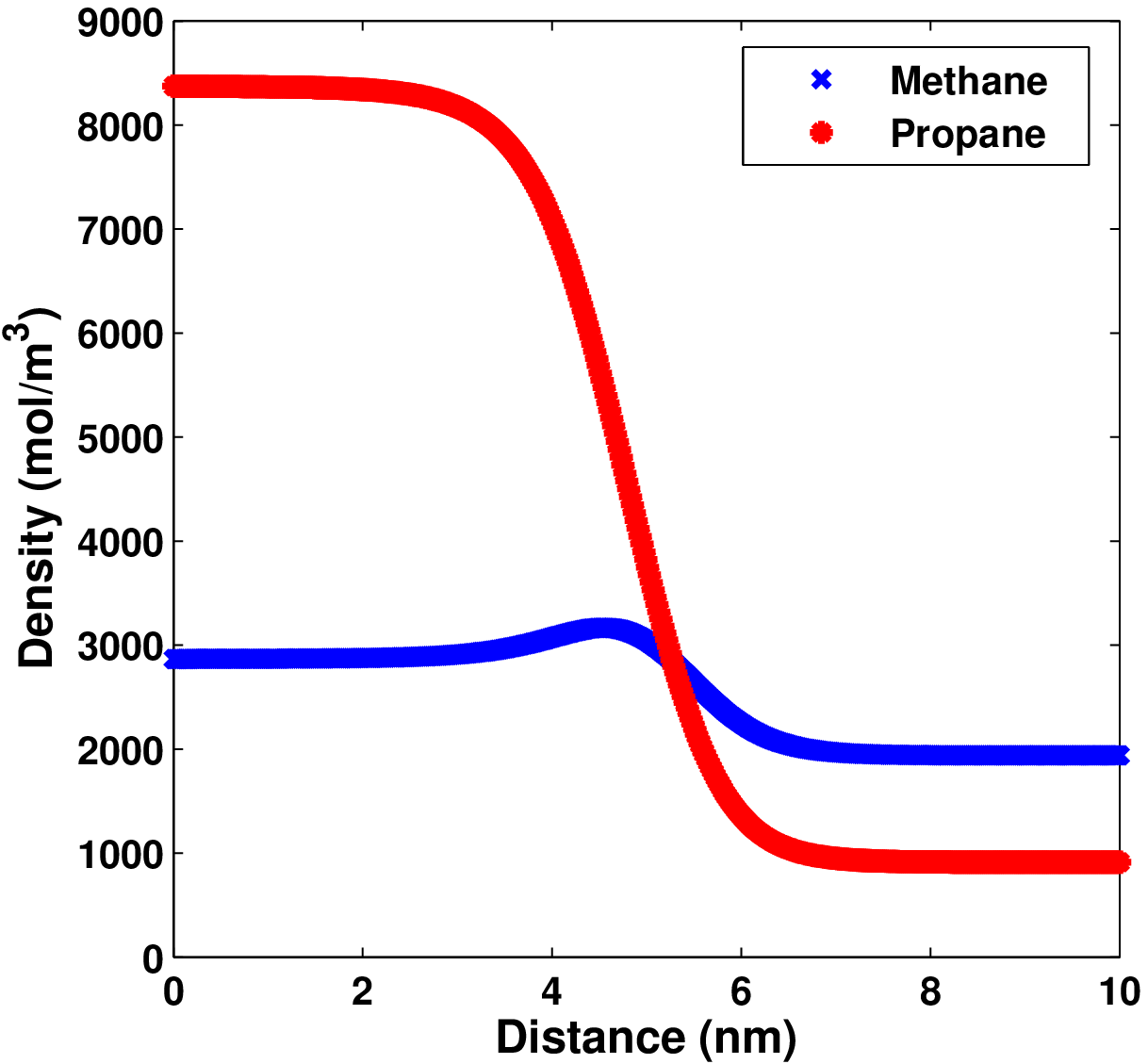}
      \caption{$s=30$.}
    \end{subfigure}
    \caption[Calculated density profile of hexane from a random initial guess.]%
            {Numerical stability test: use random density distributions as initial guess (methane--propane, 303.15~K, 60~bar).}
    \label{fig:randommixture}
  \end{figure}

   
  In the numerical stability test, random density distributions are used as initial condition to start the calculation. Densities on each local point are produced by a random number generator bounded by bulk densities, as shown in Figure~\ref{fig:randommixture}a. The density profile successfully evolves to an equilibrium state with a time step of $\Delta s = 0.1$. Based on the equilibrium density distribution, the surface tension is calculated and compared with experimental data in Table. \ref{tab:numericaltestmixture}. A very good agreement is obtained with AD~0.129\%. 
  %It is important to note that, compared with Figure~\ref{fig:randommixture}a, methane (cross marks) density points expand to a wider range in Figure~\ref{fig:randommixture}b, due to not bounding densities during the solving process in order to catch any possible density accumulations in the interface. 
  %For example in this test, methane accumulates in the interface region where its densities are higher than that in both bulk phases as shown in Figure~\ref{fig:randommixture}f.
  
  Another numerical experiment is conducting the calculations on different domain sizes. Robust convergence results are obtained for methane--propane system on a domain size that is close to the interface width (10~nm) as well as on a much wider domain (25~nm). Surface tension results are documented in Table \ref{tab:numericaltestmixture} which have a deviation to the experimental data of less than~0.2\%.
  
  %%%%%%%%%%%%%%%%%%%%%%%%%%%%%%%%%%%%%%%%%%%%%%%%%%%%%%%%%%%%%
  \section{Conclusion}
  %%%%%%%%%%%%%%%%%%%%%%%%%%%%%%%%%%%%%%%%%%%%%%%%%%%%%%%%%%%%%
  In this paper, a~stabilized density gradient theory algorithm (SDGT) is introduced to solve for interfacial properties of pure and mixed systems. \mbox{PC-SAFT} equation of state is employed which accurately describes phase equilibrium for a variety of mixtures. Compared with the conventional reference fluid algorithm, the SDGT has the advantages that no reference fluid is required and no estimation of the interface thickness is required. The physical performance of this algorithm is investigated by running interfacial property calculations and comparing the results with reported modeling and experimental data in several pure and mixed systems. Remarkable agreements are achieved in a~wide temperature and pressure range.
  The numerical stability is demonstrated using several extreme conditions like random initial conditions and overestimated domain sizes, which are potential risks of failing calculation in practical applications. The robustness and expandability of this stabilized algorithm is a proof for its practical utility to a~wider application of density gradient theory.  As next upcoming step, interfacial properties of mixtures with associating components (water for example) will be calculated by using stabilized density gradient theory with \mbox{PC-SAFT} EoS.
  
  %%%%%%%%%%%%%%%%%%%%%%%%%%%%%%%%%%%%%%%%%%%%%%%%%%%%%%%%%%%%%
  \section*{Acknowledgement}
  %%%%%%%%%%%%%%%%%%%%%%%%%%%%%%%%%%%%%%%%%%%%%%%%%%%%%%%%%%%%%
    The authors would like to thank Shell International Exploration and Production Inc.\ for the financial support. The authors would also like to thank $\text{Birol Dindoruk}$ for valuable discussions.
  
  %%%%%%%%%%%%%%%%%%%%%%%%%%%%%%%%%%%%%%%%%%%%%%%%%%%%%%%%%%%%%
  \section*{List of symbols}
  %%%%%%%%%%%%%%%%%%%%%%%%%%%%%%%%%%%%%%%%%%%%%%%%%%%%%%%%%%%%%
  \noindent
  \begin{supertabular}[ht!]{@{}p{25pt} p{35pt} p{200pt}}
    \textbf{Symbol}       &\textbf{Units} & \textbf{Description}\\
    % A
    $A_0$                   & \si{\joule}    & Homogeneous Helmholtz free energy\\
    $A_0^\mathrm{id}$       & \si{\joule}    & Ideal gas contribution to $A_0$\\
    $A_0^\mathrm{hs}$       & \si{\joule}    & Hard sphere contribution to $A_0$\\
    $A_0^\mathrm{hc}$       & \si{\joule}    & Hard chain contribution to $A_0$\\
    $A_0^\mathrm{disp}$     & \si{\joule}    & Dispersion contribution to $A_0$\\
    $A_0^\mathrm{assoc}$    & \si{\joule}    & Association contribution to $A_0$\\
    $a_0$                 & $\si{\joule}/\si{\meter}^3$&Helmholtz free energy density\\
    % B
    $\beta_{ij}$          & 1              & Mixing factor for influence parameter\\
    % C
    % D
    $D$                   & \si{\meter}    & Domain size of calculation\\
    % E
    $\epsilon_i$          & \si{\joule}    & Depth of pair potential\\
    $\epsilon^{\mathrm{A}_i\mathrm{B}_i}$     & \si{\joule}    & Association energy\\
    % F
    % G
    % H 
    % I
    % J
    % K
    $k_{ij}$              &1              & Binary interaction parameter\\
    $\kappa^{\mathrm{A}_i\mathrm{B}_i}$       &1              & Effective association volume\\
    % L
    $L$                   & \si{\meter}    & Interface thickness\\
    % M
    $\mu_i$               & \si{\joule}/\si{\mole}&Chemical potential of comp.~$i$\\
    $\mu_{i,\mathrm{bulk}}$&\si{\joule}/\si{\mole}& Bulk chemical potential of comp.~$i$\\
    % N
    $N$                   & 1             & Number of components in system\\
    % O
    $\Omega$              &\si{\joule}    & Grand potential energy\\
    % P
    $P_0$                 &\si{\pascal}   & Bulk pressure\\
    % Q
    % R
    $\rho_i$              &$\si{\mole}/\si{\meter}^3$& Molar density of comp.~$i$\\
    $\rho_{i,\mathrm{A}}$ & $\si{\mole}/\si{\meter}^3$& Bulk density of comp.~$i$ in phase~A\\
    $\rho_\mathrm{ref}$   &$\si{\mole}/\si{\meter}^3$& Molar density of the reference fluid\\
    % S
    $\sigma_i$            & \si{\meter}    & Segment diameter of comp.~$i$\\
    % T
    $T$                   & \si{\kelvin}    & Temperature\\
    % U
    % V
    $v_i$                 & $\si{\joule}\si{\m}^5 /\si{\mole}^2$ & Influence parameter of comp.~$i$\\
    $v_{ij}$              & $\si{\joule}\si{\m}^5 /\si{\mole}^2$ & Influence parameter\\
    % W
    % X
    % Y
    % Z
    $z$                   & \si{\meter}    & Distance\\ 
  \end{supertabular}
%%%%%%%%%%%%%%%%%%%%%%%%%%%%%%%%%%%%%%%%%%%%%%%%%%%%%%%%%%%%%
\section*{Abbreviations}
%%%%%%%%%%%%%%%%%%%%%%%%%%%%%%%%%%%%%%%%%%%%%%%%%%%%%%%%%%%%%
\noindent
\begin{tabularx}{\linewidth}{@{}lL@{}}
	BVP & Boundary value problem
	\\
	DGT & Density gradient theory
	\\
	EoS & Equation of state
	\\
	LGT & Linear gradient theory
	\\
	PC-SAFT & Perturbed chain SAFT
	\\
	RF & Reference fluid
	\\
	SAFT & Statistical associating fluid theory
	\\
	SDGT & Stabilized density gradient theory
	\\
	VLE & Vapor-liquid equilibrium
\end{tabularx}
%%%%%%%%%%%%%%%%%%%%%%%%%%%%%%%%%%%%%%%%%%%%%%%%%%%%%%%%%%%%%
\section*{References}
%%%%%%%%%%%%%%%%%%%%%%%%%%%%%%%%%%%%%%%%%%%%%%%%%%%%%%%%%%%%%
%% If you have bibdatabase file and want bibtex to generate the
%% bibitems, please use
%%
  \bibliographystyle{ieeetr} 
  \bibliography{ref}

\begin{thebibliography}{10}

\bibitem{van1894thermodynamische}
J.~Van~der Waals, {\em Thermodynamische Theorie der Kapillarit{\"a}t unter
  voraussetzung stetiger Dichte{\"a}nderung}.
\newblock 1894.

\bibitem{cahn_free_1958}
J.~W. Cahn and J.~E. Hilliard, ``Free {Energy} of a {Nonuniform} {System}. {I}.
  {Interfacial} {Free} {Energy},'' {\em The Journal of Chemical Physics},
  vol.~28, pp.~258--267, Feb. 1958.

\bibitem{carey1979gradient}
B.~S. Carey, ``The gradient theory of fluid interfaces,'' 1979.

\bibitem{cornelisse_application_1993}
P.~M.~W. Cornelisse, C.~J. Peters, and J.~de~Swaan~Arons, ``Application of the
  {Peng}-{Robinson} equation of state to calculate interfacial tensions and
  profiles at vapour-liquid interfaces,'' {\em Fluid Phase Equilibria},
  vol.~82, pp.~119--129, Feb. 1993.

\bibitem{cornelisse_non-classical_1996}
P.~M.~W. Cornelisse, C.~J. Peters, and J.~de~Swaan~Arons, ``Non-classical
  interfacial tension and fluid phase behaviour,'' {\em Fluid Phase
  Equilibria}, vol.~117, pp.~312--319, Mar. 1996.

\bibitem{poser_interfacial_1981}
C.~I. Poser and I.~C. Sanchez, ``Interfacial tension theory of low and high
  molecular weight liquid mixtures,'' {\em Macromolecules}, vol.~14,
  pp.~361--370, Mar. 1981.

\bibitem{enders_calculation_1998}
S.~Enders and K.~Quitzsch, ``Calculation of {Interfacial} {Properties} of
  {Demixed} {Fluids} {Using} {Density} {Gradient} {Theory},'' {\em Langmuir},
  vol.~14, pp.~4606--4614, Aug. 1998.

\bibitem{teletzke_gradient_1982}
G.~F. Teletzke, L.~E. Scriven, and H.~T. Davis, ``Gradient theory of wetting
  transitions,'' {\em Journal of Colloid and Interface Science}, vol.~87,
  pp.~550--571, June 1982.

\bibitem{zuo_linear_1996}
Y.-X. Zuo and E.~H. Stenby, ``A {Linear} {Gradient} {Theory} {Model} for
  {Calculating} {Interfacial} {Tensions} of {Mixtures},'' {\em Journal of
  Colloid and Interface Science}, vol.~182, pp.~126--132, Sept. 1996.

\bibitem{miqueu_modelling_2003}
C.~Miqueu, B.~Mendiboure, A.~Graciaa, and J.~Lachaise, ``Modelling of the
  surface tension of pure components with the gradient theory of fluid
  interfaces: a simple and accurate expression for the influence parameters,''
  {\em Fluid Phase Equilibria}, vol.~207, pp.~225--246, May 2003.

\bibitem{miqueu_modelling_2004}
C.~Miqueu, B.~Mendiboure, C.~Graciaa, and J.~Lachaise, ``Modelling of the
  surface tension of binary and ternary mixtures with the gradient theory of
  fluid interfaces,'' {\em Fluid Phase Equilibria}, vol.~218, pp.~189--203,
  Apr. 2004.

\bibitem{miqueu_modeling_2005}
C.~Miqueu, B.~Mendiboure, A.~Graciaa, and J.~Lachaise, ``Modeling of the
  {Surface} {Tension} of {Multicomponent} {Mixtures} with the {Gradient}
  {Theory} of {Fluid} {Interfaces},'' {\em Industrial \& Engineering Chemistry
  Research}, vol.~44, pp.~3321--3329, Apr. 2005.

\bibitem{carey_semiempirical_1978}
B.~S. Carey, L.~E. Scriven, and H.~T. Davis, ``Semiempirical theory of surface
  tensions of pure normal alkanes and alcohols,'' {\em AIChE Journal}, vol.~24,
  pp.~1076--1080, Nov. 1978.

\bibitem{carey_semiempirical_1980}
B.~S. Carey, L.~E. Scriven, and H.~T. Davis, ``Semiempirical theory of surface
  tension of binary systems,'' {\em AIChE Journal}, vol.~26, pp.~705--711,
  Sept. 1980.

\bibitem{sahimi_thermodynamic_1985}
M.~Sahimi, H.~T. Davis, and L.~E. Scriven, ``Thermodynamic {Modeling} of
  {Phase} and {Tension} {Behavior} of {Co}/{Sub} 2//{Hydrocarbon} {Systems},''
  {\em SPEJ, Soc. Pet. Eng. J.; (United States)}, vol.~25:2, Apr. 1985.

\bibitem{sahimi_surface_1991}
M.~Sahimi and B.~N. Taylor, ``Surface tension of binary liquid–vapor
  mixtures: {A} comparison of mean‐field and scaling theories,'' {\em The
  Journal of Chemical Physics}, vol.~95, pp.~6749--6761, Nov. 1991.

\bibitem{lin2007gradient}
H.~Lin, Y.-Y. Duan, and Q.~Min, ``Gradient theory modeling of surface tension
  for pure fluids and binary mixtures,'' {\em Fluid Phase Equilibria},
  vol.~254, no.~1, pp.~75--90, 2007.

\bibitem{chapman1988phase}
W.~G. Chapman, G.~Jackson, and K.~E. Gubbins, ``Phase equilibria of associating
  fluids: chain molecules with multiple bonding sites,'' {\em Molecular
  Physics}, vol.~65, no.~5, pp.~1057--1079, 1988.

\bibitem{chapman1989saft}
W.~G. Chapman, K.~E. Gubbins, G.~Jackson, and M.~Radosz, ``Saft:
  Equation-of-state solution model for associating fluids,'' {\em Fluid Phase
  Equilibria}, vol.~52, pp.~31--38, 1989.

\bibitem{chapman1990new}
W.~G. Chapman, K.~E. Gubbins, G.~Jackson, and M.~Radosz, ``New reference
  equation of state for associating liquids,'' {\em Industrial \& Engineering
  Chemistry Research}, vol.~29, no.~8, pp.~1709--1721, 1990.

\bibitem{wertheim1984fluids}
M.~Wertheim, ``Fluids with highly directional attractive forces. i. statistical
  thermodynamics,'' {\em Journal of statistical physics}, vol.~35, no.~1-2,
  pp.~19--34, 1984.

\bibitem{wertheim1984fluids2}
M.~Wertheim, ``Fluids with highly directional attractive forces. ii.
  thermodynamic perturbation theory and integral equations,'' {\em Journal of
  statistical physics}, vol.~35, no.~1-2, pp.~35--47, 1984.

\bibitem{wertheim1986fluids}
M.~Wertheim, ``Fluids with highly directional attractive forces. iii. multiple
  attraction sites,'' {\em Journal of statistical physics}, vol.~42, no.~3-4,
  pp.~459--476, 1986.

\bibitem{wertheim1986fluids2}
M.~Wertheim, ``Fluids with highly directional attractive forces. iv.
  equilibrium polymerization,'' {\em Journal of statistical physics}, vol.~42,
  no.~3-4, pp.~477--492, 1986.

\bibitem{gross2001perturbed}
J.~Gross and G.~Sadowski, ``Perturbed-chain saft: An equation of state based on
  a perturbation theory for chain molecules,'' {\em Industrial \& engineering
  chemistry research}, vol.~40, no.~4, pp.~1244--1260, 2001.

\bibitem{gross2002application}
J.~Gross and G.~Sadowski, ``Application of the perturbed-chain saft equation of
  state to associating systems,'' {\em Industrial \& engineering chemistry
  research}, vol.~41, no.~22, pp.~5510--5515, 2002.

\bibitem{baidakov1982surface}
V.~Baidakov, K.~Khvostov, and G.~Muratov, ``Surface-tension of nitrogen, oxygen
  and methane in a wide temperature-range,'' {\em ZHURNAL FIZICHESKOI KHIMII},
  vol.~56, no.~4, pp.~814--817, 1982.

\bibitem{baidakov1985surface}
V.~Baidakov and I.~Sulla, ``Surface propane and isobutane tension at
  temperatures close to critical,'' {\em Zhurnal Fizicheskoi Khimii}, vol.~59,
  no.~4, pp.~955--957, 1985.

\bibitem{grigoryev1992surface}
B.~Grigoryev, B.~Nemzer, D.~Kurumov, and J.~Sengers, ``Surface tension of
  normal pentane, hexane, heptane, and octane,'' {\em International journal of
  thermophysics}, vol.~13, no.~3, pp.~453--464, 1992.

\bibitem{kalbassi1988surface}
M.~Kalbassi and M.~Biddulph, ``Surface tensions of mixtures at their boiling
  points,'' {\em Journal of Chemical and Engineering Data}, vol.~33, no.~4,
  pp.~473--476, 1988.

\bibitem{barker1967perturbation}
J.~A. Barker and D.~Henderson, ``Perturbation theory and equation of state for
  fluids. ii. a successful theory of liquids,'' {\em The Journal of Chemical
  Physics}, vol.~47, no.~11, pp.~4714--4721, 1967.

\bibitem{wolbach1998using}
J.~P. Wolbach and S.~I. Sandler, ``Using molecular orbital calculations to
  describe the phase behavior of cross-associating mixtures,'' {\em Industrial
  \& engineering chemistry research}, vol.~37, no.~8, pp.~2917--2928, 1998.

\bibitem{qiao2014two}
Z.~Qiao and S.~Sun, ``Two-phase fluid simulation using a diffuse interface
  model with peng--robinson equation of state,'' {\em SIAM Journal on
  Scientific Computing}, vol.~36, no.~4, pp.~B708--B728, 2014.

\bibitem{pedersen2014phase}
K.~S. Pedersen, P.~L. Christensen, and J.~A. Shaikh, {\em Phase behavior of
  petroleum reservoir fluids}.
\newblock CRC Press, 2014.

\bibitem{weinaug1943surface}
C.~F. Weinaug and D.~L. Katz, ``Surface tensions of methane-propane mixtures,''
  {\em Industrial \& Engineering Chemistry}, vol.~35, no.~2, pp.~239--246,
  1943.

\bibitem{herdes2015coarse}
C.~Herdes, T.~S. Totton, and E.~A. M{\"u}ller, ``Coarse grained force field for
  the molecular simulation of natural gases and condensates,'' {\em Fluid Phase
  Equilibria}, vol.~406, pp.~91--100, 2015.

\bibitem{mahl1972surface}
B.~Mahl, P.~SINGH, and S.~CHOPRA, ``Surface-tension of binary-mixtures,'' {\em
  ZEITSCHRIFT FUR PHYSIKALISCHE CHEMIE-LEIPZIG}, vol.~249, no.~5-6, p.~337,
  1972.

\end{thebibliography}


\begin{thebibliography}{10}

\bibitem{rowlinson1979translation}
J.~Rowlinson, ``Translation of jd van der waals ''the thermodynamik theory of
  capillarity under the hypothesis of a continuous variation of densit'',''
  {\em Journal of Statistical Physics}, vol.~20, no.~2, pp.~197--200, 1979.

\bibitem{cahn_free_1958}
J.~W. Cahn and J.~E. Hilliard, ``Free {energy} of a {nonuniform} {system}. {I}.
  {Interfacial} {free} {energy},'' {\em The Journal of Chemical Physics},
  vol.~28, pp.~258--267, Feb. 1958.

\bibitem{carey1979gradient}
B.~S. Carey, {\em The gradient theory of fluid interfaces}.
\newblock PhD thesis, University of Minnesota, 1979.

\bibitem{cornelisse_application_1993}
P.~M.~W. Cornelisse, C.~J. Peters, and J.~de~Swaan~Arons, ``Application of the
  {Peng}--{Robinson} equation of state to calculate interfacial tensions and
  profiles at vapour-liquid interfaces,'' {\em Fluid Phase Equilibria},
  vol.~82, pp.~119--129, Feb. 1993.

\bibitem{cornelisse_non-classical_1996}
P.~M.~W. Cornelisse, C.~J. Peters, and J.~de~Swaan~Arons, ``Non-classical
  interfacial tension and fluid phase behaviour,'' {\em Fluid Phase
  Equilibria}, vol.~117, pp.~312--319, Mar. 1996.

\bibitem{poser_interfacial_1981}
C.~I. Poser and I.~C. Sanchez, ``Interfacial tension theory of low and high
  molecular weight liquid mixtures,'' {\em Macromolecules}, vol.~14,
  pp.~361--370, Mar. 1981.

\bibitem{enders_calculation_1998}
S.~Enders and K.~Quitzsch, ``Calculation of {interfacial} {properties} of
  {demixed} {fluids} {using} {density} {gradient} {theory},'' {\em Langmuir},
  vol.~14, pp.~4606--4614, Aug. 1998.

\bibitem{teletzke_gradient_1982}
G.~F. Teletzke, L.~E. Scriven, and H.~T. Davis, ``Gradient theory of wetting
  transitions,'' {\em Journal of Colloid and Interface Science}, vol.~87,
  pp.~550--571, June 1982.

\bibitem{zuo_linear_1996}
Y.-X. Zuo and E.~H. Stenby, ``A {linear} {gradient} {theory} {model} for
  {calculating} {interfacial} {tensions} of {mixtures},'' {\em Journal of
  Colloid and Interface Science}, vol.~182, pp.~126--132, Sept. 1996.

\bibitem{miqueu_modelling_2003}
C.~Miqueu, B.~Mendiboure, A.~Graciaa, and J.~Lachaise, ``Modelling of the
  surface tension of pure components with the gradient theory of fluid
  interfaces: a simple and accurate expression for the influence parameters,''
  {\em Fluid Phase Equilibria}, vol.~207, pp.~225--246, May 2003.

\bibitem{miqueu_modelling_2004}
C.~Miqueu, B.~Mendiboure, C.~Graciaa, and J.~Lachaise, ``Modelling of the
  surface tension of binary and ternary mixtures with the gradient theory of
  fluid interfaces,'' {\em Fluid Phase Equilibria}, vol.~218, pp.~189--203,
  Apr. 2004.

\bibitem{miqueu_modeling_2005}
C.~Miqueu, B.~Mendiboure, A.~Graciaa, and J.~Lachaise, ``Modeling of the
  surface tension of multicomponent mixtures with the gradient theory of fluid
  interfaces,'' {\em Industrial \& Engineering Chemistry Research}, vol.~44,
  pp.~3321--3329, Apr. 2005.

\bibitem{kou2015efficient}
J.~Kou, S.~Sun, and X.~Wang, ``Efficient numerical methods for simulating
  surface tension of multi-component mixtures with the gradient theory of fluid
  interfaces,'' {\em Computer Methods in Applied Mechanics and Engineering},
  vol.~292, pp.~92--106, 2015.

\bibitem{carey_semiempirical_1978}
B.~S. Carey, L.~E. Scriven, and H.~T. Davis, ``Semiempirical theory of surface
  tensions of pure normal alkanes and alcohols,'' {\em AIChE Journal}, vol.~24,
  pp.~1076--1080, Nov. 1978.

\bibitem{carey_semiempirical_1980}
B.~S. Carey, L.~E. Scriven, and H.~T. Davis, ``Semiempirical theory of surface
  tension of binary systems,'' {\em AIChE Journal}, vol.~26, pp.~705--711,
  Sept. 1980.

\bibitem{sahimi_thermodynamic_1985}
M.~Sahimi, H.~T. Davis, and L.~E. Scriven, ``Thermodynamic {modeling} of
  {phase} and {tension} {behavior} of {CO}/{sub} 2//{Hydrocarbon} {Systems},''
  {\em SPEJ, Soc. Pet. Eng. J.; (United States)}, vol.~25:2, Apr. 1985.

\bibitem{sahimi_surface_1991}
M.~Sahimi and B.~N. Taylor, ``Surface tension of binary liquid--vapor mixtures:
  A comparison of mean-field and scaling theories,'' {\em The Journal of
  Chemical Physics}, vol.~95, pp.~6749--6761, Nov. 1991.

\bibitem{lin2007gradient}
H.~Lin, Y.-Y. Duan, and Q.~Min, ``Gradient theory modeling of surface tension
  for pure fluids and binary mixtures,'' {\em Fluid Phase Equilibria},
  vol.~254, no.~1, pp.~75--90, 2007.

\bibitem{chapman1988phase}
W.~G. Chapman, G.~Jackson, and K.~E. Gubbins, ``Phase equilibria of associating
  fluids: chain molecules with multiple bonding sites,'' {\em Molecular
  Physics}, vol.~65, no.~5, pp.~1057--1079, 1988.

\bibitem{chapman1989saft}
W.~G. Chapman, K.~E. Gubbins, G.~Jackson, and M.~Radosz, ``Saft:
  Equation-of-state solution model for associating fluids,'' {\em Fluid Phase
  Equilibria}, vol.~52, pp.~31--38, 1989.

\bibitem{chapman1990new}
W.~G. Chapman, K.~E. Gubbins, G.~Jackson, and M.~Radosz, ``New reference
  equation of state for associating liquids,'' {\em Industrial \& Engineering
  Chemistry Research}, vol.~29, no.~8, pp.~1709--1721, 1990.

\bibitem{wertheim1984fluids}
M.~Wertheim, ``Fluids with highly directional attractive forces. {I}.
  statistical thermodynamics,'' {\em Journal of Statistical Physics}, vol.~35,
  no.~1-2, pp.~19--34, 1984.

\bibitem{wertheim1984fluids2}
M.~Wertheim, ``Fluids with highly directional attractive forces. {II}.
  thermodynamic perturbation theory and integral equations,'' {\em Journal of
  Statistical Physics}, vol.~35, no.~1-2, pp.~35--47, 1984.

\bibitem{wertheim1986fluids}
M.~Wertheim, ``Fluids with highly directional attractive forces. {III}.
  multiple attraction sites,'' {\em Journal of Statistical Physics}, vol.~42,
  no.~3-4, pp.~459--476, 1986.

\bibitem{wertheim1986fluids2}
M.~Wertheim, ``Fluids with highly directional attractive forces. {IV}.
  equilibrium polymerization,'' {\em Journal of Statistical Physics}, vol.~42,
  no.~3-4, pp.~477--492, 1986.

\bibitem{vega201120years}
L.~F. Vega and G.~Jackson, ``20 years of the {SAFT} equation of state--recent
  advances and challenges: Symposium held in bellaterra, barcelona, 19--21
  september 2010,'' {\em Fluid Phase Equilibria}, vol.~306, no.~1, pp.~1--3,
  2011.

\bibitem{gross2001perturbed}
J.~Gross and G.~Sadowski, ``Perturbed-chain {SAFT}: An equation of state based
  on a perturbation theory for chain molecules,'' {\em Industrial \&
  Engineering Chemistry Research}, vol.~40, no.~4, pp.~1244--1260, 2001.

\bibitem{gross2002application}
J.~Gross and G.~Sadowski, ``Application of the perturbed-chain {SAFT} equation
  of state to associating systems,'' {\em Industrial \& Engineering Chemistry
  Research}, vol.~41, no.~22, pp.~5510--5515, 2002.

\bibitem{Baidakov1982}
V.~Baidakov, K.~Khvostov, and G.~Muratov, ``Surface-tension of nitrogen, oxygen
  and methane in a wide temperature-range,'' {\em Zhurnal Fizicheskoi Khimii},
  vol.~56, no.~4, pp.~814--817, 1982.

\bibitem{baidakov1985surface}
V.~Baidakov and I.~Sulla, ``Surface propane and isobutane tension at
  temperatures close to critical,'' {\em Zhurnal Fizicheskoi Khimii}, vol.~59,
  no.~4, pp.~955--957, 1985.

\bibitem{Grigoryev1992}
B.~Grigoryev, B.~Nemzer, D.~Kurumov, and J.~Sengers, ``Surface tension of
  normal pentane, hexane, heptane, and octane,'' {\em International Journal of
  Thermophysics}, vol.~13, no.~3, pp.~453--464, 1992.

\bibitem{kalbassi1988surface}
M.~Kalbassi and M.~Biddulph, ``Surface tensions of mixtures at their boiling
  points,'' {\em Journal of Chemical and Engineering Data}, vol.~33, no.~4,
  pp.~473--476, 1988.

\bibitem{barker1967perturbation}
J.~A. Barker and D.~Henderson, ``Perturbation theory and equation of state for
  fluids. ii. a successful theory of liquids,'' {\em The Journal of Chemical
  Physics}, vol.~47, no.~11, pp.~4714--4721, 1967.

\bibitem{wolbach1998using}
J.~P. Wolbach and S.~I. Sandler, ``Using molecular orbital calculations to
  describe the phase behavior of cross-associating mixtures,'' {\em Industrial
  \& Engineering Chemistry Research}, vol.~37, no.~8, pp.~2917--2928, 1998.

\bibitem{qiao2014two}
Z.~Qiao and S.~Sun, ``Two-phase fluid simulation using a diffuse interface
  model with {Peng}--{Robinson} equation of state,'' {\em SIAM Journal on
  Scientific Computing}, vol.~36, no.~4, pp.~B708--B728, 2014.

\bibitem{ElliottStuart1993}
C.~M. Elliott and A.~M. Stuart, ``The global dynamics of discrete semilinear
  parabolic equations,'' {\em SIAM J. Numer. Anal.}, vol.~30, no.~6,
  pp.~1622--1663, 1993.

\bibitem{eyre1998unconditionally}
D.~J. Eyre, ``An unconditionally stable one-step scheme for gradient systems,''
  {\em Unpublished article}, 1998.

\bibitem{pedersen2014phase}
K.~S. Pedersen, P.~L. Christensen, and J.~A. Shaikh, {\em Phase behavior of
  petroleum reservoir fluids}.
\newblock CRC Press, 2014.

\bibitem{weinaug1943surface}
C.~F. Weinaug and D.~L. Katz, ``Surface tensions of methane-propane mixtures,''
  {\em Industrial \& Engineering Chemistry}, vol.~35, no.~2, pp.~239--246,
  1943.

\bibitem{rowlinson2013molecular}
J.~S. Rowlinson and B.~Widom, {\em Molecular Theory of Capillarity}.
\newblock Courier Corporation, 2013.

\bibitem{Sergio19th}
{{Sergio} {E.} {Qui{\~n}ones}-{Cisneros}, {Eder} {L.} {Granados}-{Baz{\'a}n}
  and {Ulrich} {K.} {Deiters}}, ``Estimation of multicomponent interfacial
  density profiles from a direct minimization of the free energy surface,'' in
  {\em Nineteenth Symposium on Thermophysical Properties, Boulder, CO, USA},
  2015.

\bibitem{herdes2015coarse}
C.~Herdes, T.~S. Totton, and E.~A. M{\"u}ller, ``Coarse grained force field for
  the molecular simulation of natural gases and condensates,'' {\em Fluid Phase
  Equilibria}, vol.~406, pp.~91--100, 2015.

\bibitem{Mahl1972}
B.~Mahl, P.~Singh, and S.~Chopra, ``Surface-tension of binary-mixtures,'' {\em
  Zeitschrift f\"ur physikalische Chemie Leipzig}, vol.~249, no.~5-6, p.~337,
  1972.

\end{thebibliography}

%% else use the following coding to input the bibitems directly in the
%% TeX file.

%\begin{thebibliography}{00}

%% \bibitem{label}
%% Text of bibliographic item

%\bibitem{}

%\end{thebibliography}
\end{document}